\documentclass[dvips,12pt] {article}
\usepackage{epsfig}
\usepackage{a4wide}

\usepackage{amsmath}
\usepackage{latexsym}
\usepackage{multirow}
\newcommand{\dd}{\mbox{{\rm d}}}
\newcommand{\half}{\textstyle\frac{1}{2}}
\newcommand{\wmax}{w_{\rm max}}
\renewcommand{\Re}{\mbox{Re}}
\def\deg{\hbox{$^{\circ}$}}
\newcommand{\thW}{\theta_{{\rm W}}}
\newcommand{\varepsilonmax}{\varepsilon_{\rm max}}

\newcommand{\veck}{{\pmb{k}}}
\newcommand{\vecp}{{\pmb{p}}}
\def\Month{\ifcase\month\or
January\or February\or March\or April\or May\or June\or
July\or August\or September\or October\or November\or December\fi}
\begin{document}
\thispagestyle{empty}
\begin{flushright}
{\tt University of Bergen, Department of Physics}    \\[2mm]
{\tt Scientific/Technical Report No.1998-09}    \\[2mm]
{\tt ISSN 0803-2696} \\[5mm]
{hep-ph/9811505} \\[5mm]
{November 1998}           \\
\end{flushright}

\vspace*{3cm}
\begin{center}
{\bf \Large
Higgs Production: $CP$ Studies in $e^-e^-$ Collisions}
\end{center}
\vspace{1cm}
\begin{center}
C.~A.\ B{\o}e$^{a}$, O.~M.\ Ogreid$^{a}$, P.~Osland$^{a}$ 
\ and \
Jian-zu Zhang$^{b}$ \\
\vspace{1cm}
$^{a}${\em
Department of Physics, University of Bergen, \\
      All\'{e}gaten 55, N-5007 Bergen, Norway}
\\
\vspace{.3cm}
$^{b}${\em
School of Science,
East China University of Science and Technology, \\
130 Mei Long Road,
Shanghai 200237, P. R. China}
\end{center}
\hspace{3in}

\begin{abstract}
We review the production of scalar Higgs-like particles in high-energy
electron-electron collisions,
via the fusion of electroweak gauge bosons.
The emphasis is on how to distinguish a
$CP$-even from a $CP$-odd Higgs particle.
Among the more significant differences, we find that
in the $CP$-odd case, the Higgs spectrum is much harder,
and the dependence of the total cross section on the product of the
polarizations of the two beams is much stronger, than in
the $CP$-even case.
We also briefly discuss parity violation, and the production of
charged Higgs bosons.
\end{abstract}
\vfil

%\pagestyle{myheadings}
%\markboth{\Draft{Draft}}{\Draft{Draft}}

%%%%%%%%%%%%%%%%%%%%%%%%%%%%%%%%%%%%%%%%%%%%%%%%%%%%%%%%%%%%%%%%%%%%%%%%
\section{Introduction}
\setcounter{equation}{0}
%%%%%%%%%%%%%%%%%%%%%%%%%%%%%%%%%%%%%%%%%%%%%%%%%%%%%%%%%%%%%%%%%%%%%%%%
In the planning for a future linear collider \cite{NLC,Accomando} 
one has to explore
not only the electron-positron mode and various photon modes, 
but also an electron-electron 
mode, in spite of concerns related to beam ``disruption''.
One reason for an electron-electron collider to be interesting is that one
may produce states not accessible in the annihilation channel,
another is that a large electron polarization will be readily available.
There is already a considerable literature on the electron-electron
mode \cite{Hikasa,e-e-,Gunion,Barger}.

We here consider the production of Higgs particles\footnote{The
term ``Higgs particle'' will here be used quite generally about
any scalar, electrically neutral or charged, that has a significant
coupling to electroweak gauge bosons.}
in electron-electron collisions.
Apart from a precise determination
of the Higgs mass, which will allow for certain consistency tests 
of the theory,
one will want to determine its properties under the discrete symmetries,
and the couplings to various other particles.

At high energies, the Higgs production at an electron-electron
collider will proceed via gauge boson fusion \cite{Hikasa,Barger}, 
and thus not be suppressed by the $s$-channel annihilation mechanism
\cite{Bjorken}.
Certain models also predict doubly charged Higgs particles \cite{double-h},
some of which can be produced more readily at an electron-electron collider.

Scalar (``Higgs'') particles, $h$, $h^-$ and $h^{--}$, 
are produced in the $t$-channel via $Z$- or $W$-exchange:
\begin{eqnarray}
\label{eq-e-e-}
e^-(p_1)+e^-(p_2) &\rightarrow& e^-(p'_1)+e^-(p'_2)+h(p_h), \\
e^-(p_1)+e^-(p_2) &\rightarrow& e^-(p'_1)+\nu_e(p'_2)+h^-(p_h), \\
e^-(p_1)+e^-(p_2) &\rightarrow& \nu_e(p'_1)+\nu_e(p'_2)+h^{--}(p_h),
\end{eqnarray}
as depicted in Fig.~\ref{Fig:feynman} for the case (\ref{eq-e-e-}).
(However, in some models, including the left--right symmetric model 
\cite{Mohapatra}, the doubly-charged Higgs boson has practically
no coupling to the ordinary, left-handed $W$ bosons. They would not
be produced by this mechanism.)

It is well known that the $CP$ property of the Higgs particle can 
be explored in the electron-positron annihilation mode 
from studies of angular and energy correlations \cite{e-p,SkjOsl95,Arens}.
In the present paper we analyze the corresponding situation for
the $t$-channel, at an electron-electron collider, 
taking into account the effects of beam polarization.  

We shall investigate to what extent various angular distributions
and energy correlations are sensitive to whether the Higgs particle
is even or odd under $CP$, in which case it will be denoted as
$H$ or $A$, respectively. 
It turns out that several of these distributions
are quite sensitive to the $CP$ property of the Higgs particle.
Some of these results were presented elsewhere \cite{Ogreid}.

The $ZZh$ coupling is taken to be \cite{NelCha}
\begin{equation}
\label{Eq:coupling}
i2^{5/4}\sqrt{G_F}
\begin{cases}
m_Z^2\,g^{\mu\nu} & \mbox{for }h=H \mbox{ ($CP$ even)}, \\
\eta\, \epsilon^{\mu\nu\rho\sigma} k_{1\rho} k_{2\sigma} &
\mbox{for }h=A \mbox{ ($CP$ odd)},
\end{cases}
\end{equation}
where $k_1$ and $k_2$ are the momenta of the gauge bosons.
Thus, we see immediately that near the forward direction,
where $\veck_1$ and $\veck_2$ are antiparallel,
the production of a $CP$-odd Higgs boson will be suppressed.
In the MSSM, this $ZZA$ coupling is absent at the tree level,
but will be induced at the 1-loop level \cite{HHG}.
Our analysis is not restricted to any particular model.

There could also be $CP$ violation in the Higgs sector, 
in which case the Higgs bosons would not be $CP$ eigenstates 
\cite{Bernreuther}. 
Such mixing could take place at the tree level
\cite{Deshpande}, or it could be induced by radiative corrections.
It has also been pointed out that such mixing might
take place in the MSSM, and be resonant \cite{Pilaftsis}.
We shall discuss ways to look for $CP$ violation in the effective
$ZZh$ coupling.

The focus will be on a light Higgs boson, as is favored by current 
LEP precision data \cite{LEP}, and the case of $E_{\rm cm}=500$~GeV
\cite{Accomando}.

When one or both $Z$'s are replaced by $W$'s (for the production
of charged Higgs particles), we shall assume that 
the Lorentz structure of the coupling remains unchanged.

The paper is organized as follows.
In Sect.~2 we give notations and discuss kinematics, in Sect.~3
we give various cross section formulas.
In Sects.\ 4--6 we present a variety of numerical results:
Section~4 is devoted to integrated cross sections and distributions 
where the final-state electrons are integrated over.
These distributions would qualitatively be the same for the production
of charged Higgs particles.
In Sect.~5 we study correlations between the final-state electrons,
and in Sect.~6 we consider parity violation.
Sect.~7 is devoted to a brief, qualitative, discussion of charged
Higgs particles, and in Sect.~8 we discuss statistics, with some
concluding remarks in Sect.~9.
%%%%%%%%%%%%%%%%%%%%%%%%%%%%%%%%%%%%%%%%%%%%%%%%%%%%%%%%%%%%%%%%%%%%%%
\section{Notation and kinematics}
\setcounter{equation}{0}
%%%%%%%%%%%%%%%%%%%%%%%%%%%%%%%%%%%%%%%%%%%%%%%%%%%%%%%%%%%%%%%%%%%%%%
The $eeZ$ vector and axial vector couplings are denoted
$g_V$ and $g_A$, as defined by the interaction
$\overline\psi(x)\gamma^\mu(g_V-g_A\gamma_5)\psi(x)Z_\mu(x)$.
As a parameterization of their ratio, we define the angle $\chi$ by
\begin{equation}
g_{V} \equiv \tilde g \cos \chi, \qquad
g_{A} \equiv \tilde g \sin \chi,
\end{equation}
with
\begin{equation}
\tilde g^2 =\left(\frac{g}{4\cos\thW}\right)^2[(1-4\sin^2\thW)^2+1],
\end{equation}
and $g$ the SU(2) electroweak coupling constant.
In the present work, the only reference to this angle $\chi$
is through $\sin2\chi$. In the case of the $eeZ$ coupling,
we have $\sin2\chi\simeq 0.1393$, whereas for the $e\nu W$ coupling,
which is purely left-handed, we have $\sin2\chi=1$.

The momenta of the final-state leptons will be referred 
to by polar angles $\theta_1$ and $\theta_2$ [see Eq.~(\ref{eq-e-e-})], 
and that of the Higgs particle by the polar
angle $\theta_h$:
\begin{eqnarray}
\label{Eq:polar-angles-e+e-}
\vecp_1\cdot\vecp_1'
& = & |\vecp_1||\vecp_1'|\cos\theta_1 
= EE_1'\cos\theta_1, \nonumber \\
\vecp_1\cdot\vecp_2'
& = & |\vecp_1||\vecp_2'|\cos\theta_2 
= EE_2'\cos\theta_2, \nonumber \\
\vecp_1\cdot\vecp_h
& = & |\vecp_1||\vecp_h|\cos\theta_h
= E\sqrt{E_h^2-m_h^2}\cos\theta_h.
\end{eqnarray}
For ``forward'' production, we will thus have
$\cos\theta_1\simeq1$, $\cos\theta_2\simeq-1$.
Furthermore, an azimuthal angle
$\phi$ will refer to the relative orientation of the two planes
formed by the final and initial-state leptons
(in ref.~\cite{Ogreid} the definition used was $\cos\phi\to-\cos\phi$),
\begin{equation}
\label{Eq:phi}
\cos{\phi}=\frac{(\vecp_1\times\vecp_1^\prime)
\cdot(\vecp_1\times\vecp_2^\prime)}
{|\vecp_1\times\vecp_1^\prime|
|\vecp_1\times\vecp_2^\prime|} \,.
\end{equation}
The two beams will be taken to be longitudinally polarized, 
with degrees of polarizations given by $P_1$ and $P_2$, respectively
($P_i>0$ for a right-handed polarization).

We shall express the cross sections in terms of the variables
\begin{eqnarray}
s_1 &=& (p_1+p_2)^2,  \qquad
s_2  =  (p_1'+p_2')^2, \nonumber \\
t_1 &=& (p_1-p_1')^2, \qquad 
t_2  =  (p_2-p_2')^2, \nonumber \\
u_1 &=& (p_1-p_2')^2, \qquad
u_2  =  (p_2-p_1')^2, 
\end{eqnarray}
where (neglecting the electron mass)
\begin{equation}
m_h^2=s_1+s_2+t_1+t_2+u_1+u_2.
\end{equation}
For the two final-state electrons, we distinguish $p_1'$ 
and $p_2'$, according to which has the higher energy, $E_1'>E_2'$.

%%%%%%%%%%%%%%%%%%%%%%%%%%%%%%%%%%%%%%%%%%%%%%%%%%%%%%%%%%%%%%%%%%%%%%
\section{The $e^-e^-\to h\, e^-e^-$ cross section}
\setcounter{equation}{0}
%%%%%%%%%%%%%%%%%%%%%%%%%%%%%%%%%%%%%%%%%%%%%%%%%%%%%%%%%%%%%%%%%%%%%%
For Higgs production from an electron-electron initial state,
via the so-called fusion mechanism (with $Z$ exchange),
there are two diagrams, because of the symmetry of the two electrons
in the final state.
The corresponding two amplitudes differ by the substitutions 
$p_1'\leftrightarrow p_2'$, corresponding to
$(t_1,t_2)\leftrightarrow(u_1,u_2)$,
and by an over-all sign.

We present the differential cross section in two different forms.
Both are useful, according to which distribution we want to study.
For the study of distributions of the final-state electrons,
we express the differential cross section as
\begin{eqnarray}
\label{EQ:1-dsigma}
\frac{\dd^4\sigma^{(h)}}{\dd\varepsilon\, 
                     \dd\cos\theta_1\, \dd\cos\theta_2\, \dd\phi}
&=&C^{(h)}\bigl\{|F(t_1,t_2)|^2\, X^{(h)} 
+ (t_j\leftrightarrow u_j) \nonumber \\
&& \phantom{C^{(h)}\bigl\{}
+ 2 \Re [F^*(u_1,u_2) F(t_1,t_2)] Z^{(h)}\bigr\},
\end{eqnarray}
where $F(t_1,t_2)$ is a propagator factor,
\begin{equation}
\label{eq-2-F}
F(t_1,t_2)=\frac{1}{t_1-m_Z^2}\, \frac{1}{t_2-m_Z^2}.
\end{equation}
The over-all constant is given as\footnote{The normalizations 
of $C^{(h)}$ and $\epsilon$ take
into account the fact that there are two identical particles
in the final state. Since the two electrons are distinguished by
their energies, the polar angles $\theta_1$ and $\theta_2$ may
take on any values in the range $[0,\pi]$.}
\begin{equation}
C^{(h)}=\frac{1}{(2\pi)^4}\frac{G_{\rm F}}{\sqrt{2}}
\frac{\tilde g^4}{2s}\frac{m_Z^4 E_1' E_2'}{|J|E_h}
\left\{
\begin{array}{ll}
1  & \mbox{for }h=H \mbox{ ($CP$ even)}, \\
\eta^2 & \mbox{for }h=A \mbox{ ($CP$ odd)}, 
\end{array}
\right.
\end{equation}
with the Jacobian 
\begin{equation}
J=1+\frac{2E-E_h}{2E_h}(1+\hat{\vecp}_1'
                     \cdot\hat{\vecp}_2'),
\end{equation}
and $\varepsilon$ half the energy difference between
the two electrons, 
\begin{equation}
\label{Eq:varepsilon}
\varepsilon=\half(E_1'-E_2').
\end{equation}
Since the two final-state electrons are indistinguishable,
we shall identify the momenta such that $E_1'\ge E_2'$;
thus, $\epsilon\ge0$.
The maximum value is given by the beam energy and the Higgs mass 
as
\begin{equation}
\label{Eq:varepsilonmax}
\varepsilonmax=\frac{1}{2}E-\frac{m_h^2}{8E}.
\end{equation}

For the purpose of studying distributions in $\cos\theta_h$
and $E_h$,
it is more convenient to express the cross section as\footnote{When 
integrating over Eq.~(\ref{EQ:1-dsigma-theta_h-E_h}) to obtain
less differential cross sections, one has to keep in mind that
there are two identical electrons in the final state, and integrate
$\cos\theta_1$ over only one hemisphere.}
\begin{eqnarray}
\label{EQ:1-dsigma-theta_h-E_h}
\frac{\dd^4\sigma^{(h)}}{\dd E_h\, 
                     \dd\cos\theta_1\, \dd\cos\theta_h\, \dd\phi_h}
&=&\tilde C^{(h)}\bigl\{|F(t_1,t_2)|^2\, X^{(h)} 
+ (t_j\leftrightarrow u_j) \nonumber \\
&& \phantom{\tilde C^{(h)}\bigl\{}
+ 2 \Re [F^*(u_1,u_2) F(t_1,t_2)] Z^{(h)}\bigr\},
\end{eqnarray}
where the over-all constant is given as 
\begin{equation}
\tilde C^{(h)}=\frac{1}{(2\pi)^4}\frac{G_{\rm F}}{\sqrt{2}}
\frac{\tilde g^4}{2s}\frac{m_Z^4 E_1'\sqrt{E_h^2-m_h^2}}{|\tilde J|E_2'}
\left\{
\begin{array}{ll}
1  & \mbox{for }h=H \mbox{ ($CP$ even)}, \\
\eta^2 & \mbox{for }h=A \mbox{ ($CP$ odd)},
\end{array}
\right.
\end{equation}
and $\tilde J$ is the Jacobian, 
\begin{equation}
\tilde J=1+\frac{1}{E_2'}\left(E_1'                     
+(\hat{\vecp}_1'\cdot\hat{\vecp}_h)
\sqrt{E_h^2-m_h^2}\right).
\end{equation}

The dynamics is given by $X^{(h)}$ and $Z^{(h)}$. 
We shall below consider three
cases: (1) The $CP$-even case, (2) the $CP$-odd case,
and (3) the case of $CP$ violation.
%%%%%%%%%%%%%%%%%%%%%%%%%%%%%%%%%%%%%%%%%%%%%%%%%%%%%%%%%%%%%%%%%%%%%%%%
\subsection{The $CP$-even case}
%%%%%%%%%%%%%%%%%%%%%%%%%%%%%%%%%%%%%%%%%%%%%%%%%%%%%%%%%%%%%%%%%%%%%%%%
For the $CP$-even case, we find
\begin{eqnarray}
\label{Eq:XZ-even}
X^{(H)}&=&
2\bigl[(1-P_1\sin2\chi)(1-P_2\sin2\chi)(s_1 s_2 +u_1 u_2) \nonumber \\
& & \phantom{2\bigl[}
+(\sin2\chi-P_1)(\sin2\chi-P_2)(s_1 s_2 -u_1 u_2)\bigr], \nonumber \\
Z^{(H)} &=&2\bigl[ (1-P_1\sin2\chi) (1-P_2\sin2\chi)
       + (\sin2\chi-P_1) (\sin2\chi-P_2) \bigr]s_1 s_2.
\end{eqnarray}
%%%%%%%%%%%%%%%%%%%%%%%%%%%%%%%%%%%%%%%%%%%%%%%%%%%%%%%%%%%%%%%%%%%%%%%%
\subsection{The $CP$-odd case}
%%%%%%%%%%%%%%%%%%%%%%%%%%%%%%%%%%%%%%%%%%%%%%%%%%%%%%%%%%%%%%%%%%%%%%%%
For the $CP$-odd case, we find
\begin{eqnarray}
\label{Eq:XZ-odd}
X^{(A)}&=&
(1-P_1\sin2\chi) (1-P_2\sin2\chi) Y_0 
+ (\sin2\chi-P_1) (\sin2\chi-P_2) Y_2 , \nonumber \\
Z^{(A)}&=&
\bigl[ (1-P_1\sin2\chi) (1-P_2\sin2\chi) 
+ (\sin2\chi-P_1) (\sin2\chi-P_2) \bigr] Y,
\end{eqnarray}
with 
\begin{eqnarray}
Y_0&=& \frac{1}{2m_Z^4}\bigl\{t_1t_2[(s_1+s_2)^2+(u_1+u_2)^2]
       -2[(s_1s_2-u_1u_2)^2+(t_1t_2)^2] \bigr\}, \nonumber \\
Y_2&=&\frac{1}{2m_Z^4} t_1t_2[(s_1-s_2)^2-(u_1-u_2)^2],
\end{eqnarray}
and
\begin{equation}
\label{Eq:XZ-odd-tu}
Y=\frac{1}{4m_Z^4}[s_1s_2(s_1^2+s_2^2) -(s_1+s_2)^2(t_1t_2+u_1u_2)
+2(t_1t_2-u_1u_2)^2].
\end{equation}

For comparison, we give in Appendix~A the corresponding results, including
beam polarization effects, as well as the $t$-channel contribution,
for the more familiar case of
\begin{equation}
\label{eq-e+e-}
e^+(p_1)+e^-(p_2) \rightarrow e^+(p'_1)+e^-(p'_2)+h(p_h).
\end{equation}

%%%%%%%%%%%%%%%%%%%%%%%%%%%%%%%%%%%%%%%%%%%%%%%%%%%%%%%%%%%%%%%%%%%%%%%%
\subsection{$CP$ violation}
%%%%%%%%%%%%%%%%%%%%%%%%%%%%%%%%%%%%%%%%%%%%%%%%%%%%%%%%%%%%%%%%%%%%%%%%
To allow for the possibility of $CP$ violation in the interaction
between electroweak gauge bosons and the Higgs, we introduce a mixing
angle $\alpha$ as follows:
\begin{equation}
{\cal M}=\cos\alpha\,{\cal M}_{\rm even}+\sin\alpha\,{\cal M}_{\rm odd}.
\end{equation}
Thus, for $\alpha=0$ or $\pi/2$, the Higgs has even or odd $CP$,
respectively, whereas for $\sin2\alpha\ne0$ the production mechanism 
violates $CP$.
This amounts to allowing for both terms, and their interference.
The suitably averaged square of the amplitude will then take
the form
\begin{eqnarray}
\sum_{\rm spin}|{\cal M}|^2
&=&\cos^2\alpha\sum_{\rm spin}|{\cal M}_{\rm even}|^2
+\sin^2\alpha\sum_{\rm spin}|{\cal M}_{\rm odd}|^2 \nonumber \\
& &+\sin2\alpha\,\Re\sum_{\rm spin}
{\cal M}_{\rm even}^\dagger {\cal M}_{\rm odd}.
\end{eqnarray}

In the notation of Eqs.~(\ref{EQ:1-dsigma}) and 
(\ref{EQ:1-dsigma-theta_h-E_h}), we get
\begin{eqnarray}
X^{(h)}&=&\cos^2\alpha\, X^{(H)}+\sin^2\alpha\, X^{(A)}
+\sin2\alpha\, \tilde X, \nonumber \\
Z^{(h)}&=&\cos^2\alpha\, Z^{(H)}+\sin^2\alpha\, Z^{(A)},
\end{eqnarray}
where the amount of $CP$ violation is given by $\sin2\alpha$, with
\begin{equation}
\tan\alpha=\eta.
\end{equation}
The $CP$-even terms $X^{(H)}$ and $Z^{(H)}$ are given by 
Eq.~(\ref{Eq:XZ-even}), and the $CP$-odd ones, $X^{(A)}$ and $Z^{(A)}$, 
by (\ref{Eq:XZ-odd})--(\ref{Eq:XZ-odd-tu}). 

There is no $CP$-violating contribution to the interference between
the $t$- and $u$-channel terms.
For the $t$- and $u$-channel $CP$-violating terms, 
which are proportional to
\begin{eqnarray}
\label{Eq:H-cp-viol}
H&\equiv&\frac{1}{m_Z^2}\,\epsilon_{\mu\nu\rho\sigma}
p_1^\mu p_2^\nu p_1'{}^\rho p_2'{}^\sigma \nonumber \\
&=&-2(E/m_Z)^2\, \hat{\vecp}_1\cdot
(\vecp'_1\times\vecp'_2),
\end{eqnarray}
we find
\begin{eqnarray}
\label{Eq:X-tu-tilde}
\tilde X_t&=& 2\{[(1+P_1P_2)(1+\sin^22\chi)-2(P_1+P_2)\sin2\chi](s_1+s_2)
\nonumber \\
& & \phantom{2\{}
-(1-P_1P_2)(1-\sin^22\chi)(u_1+u_2)\}H, \nonumber \\
\tilde X_u&=&-2\{[(1+P_1P_2)(1+\sin^22\chi)-2(P_1+P_2)\sin2\chi](s_1+s_2)
\nonumber \\
& & \phantom{-2\{}
-(1-P_1P_2)(1-\sin^22\chi)(t_1+t_2)\}H.
\end{eqnarray}

A quantitative study of this case of $CP$ violation is presented in Sec.~6.
%%%%%%%%%%%%%%%%%%%%%%%%%%%%%%%%%%%%%%%%%%%%%%%%%%%%%%%%%%%%%%%%%%%%%%%%
\section{Gross features of the cross section}
\setcounter{equation}{0}
%%%%%%%%%%%%%%%%%%%%%%%%%%%%%%%%%%%%%%%%%%%%%%%%%%%%%%%%%%%%%%%%%%%%%%%%
The $Z$ propagators will favor production at small momentum
transfers, i.e., with the final-state electrons close to the beam
directions. This is indeed how the $CP$-even Higgs particle is produced.
However, a finite momentum transfer is required to produce a $CP$-odd
particle, as is seen from the coupling (\ref{Eq:coupling}) and also
from the explicit expressions (\ref{Eq:XZ-odd})--(\ref{Eq:XZ-odd-tu}).
This statement will be illustrated quantitatively in the following.
%%%%%%%%%%%%%%%%%%%%%%%%%%%%%%%%%%%%%%%%%%%%%%%%%%%%%%%%%%%%%%%%%%%%%%%%
\subsection{Total cross section}
%%%%%%%%%%%%%%%%%%%%%%%%%%%%%%%%%%%%%%%%%%%%%%%%%%%%%%%%%%%%%%%%%%%%%%%%

For a collider at $\sqrt{s}=500$~GeV, the cross section for producing
a Standard Model Higgs with a mass of 100~GeV, is 9~fb,
and falling steeply with mass,
as illustrated in Fig.~\ref{Fig:sigtot-mh} (denoted ``even'').
The corresponding Bjorken cross section is around 60~fb \cite{Accomando}.
We also compare with the cross section for
producing a $CP$-odd Higgs boson, taking the coupling strength
$\eta$ such that the two cross sections coincide at 
$m_h=100$~GeV.\footnote{Clearly, in this phenomenological coupling, 
Eq.~(\ref{Eq:coupling}), the strength $\eta$ might depend on the Higgs mass.}

Since it may be difficult to observe electrons at small angles,
and in order to reduce certain backgrounds,
we also study the effect of a cut, with respect to the beam,
on the polar angles of the final-state electrons.
Three sets of curves are given in Fig.~\ref{Fig:sigtot-mh}, 
the upper ones are for no cut,
whereas the lower ones correspond to cuts at $5\deg$ (as suggested
by Minkowski \cite{Barger}) and at $15\deg$.\footnote{This more conservative 
cut was studied by Barger et al.\ \cite{Barger}}
Similar results are given by Hikasa \cite{Hikasa} at higher 
energies\footnote{Our cross section agrees with Fig.~6 of \cite{Hikasa}.}, 
and in \cite{Barger}.

The energy dependence is illustrated in Fig.~\ref{Fig:sigtot-e}.
As the energy increases, the cross section grows.
This is characteristic of the $t$-channel fusion mechanism,
and rather different from the case of the Bjorken mechanism.
However, an angular cut will temper this growth with energy;
for the $CP$-even case, the cross section may even decrease
with energy (see also \cite{Barger}).

Polarization-dependent total cross sections will be discussed 
in Sec.~4.4 below.
%%%%%%%%%%%%%%%%%%%%%%%%%%%%%%%%%%%%%%%%%%%%%%%%%%%%%%%%%%%%%%%%%%%%%%%%
\subsection{Higgs energy distributions}
%%%%%%%%%%%%%%%%%%%%%%%%%%%%%%%%%%%%%%%%%%%%%%%%%%%%%%%%%%%%%%%%%%%%%%%%
An interesting observable to consider, is the Higgs energy distribution
\begin{equation}
\frac{1}{\sigma^{(h)}}
\frac{\dd\sigma^{(h)}}{\dd E_h}
=\frac{1}{\sigma^{(h)}}
\int_{-1}^{1}\dd \cos\theta_1
\int_{-1}^{1}\dd \cos\theta_h
\int_0^{2\pi}\dd\phi\,
\frac{\dd^4\sigma^{(h)}}{\dd E_h \,
                     \dd\cos\theta_1\, \dd\cos\theta_h\, \dd\phi} \, .
\end{equation}
We show in Fig.~\ref{Fig:eh} such distributions,
for $\sqrt{s}=500$~GeV, and for two Higgs masses,
$m_h=120$~GeV and 150~GeV.
In the $CP$-even case, the Higgs particle is rather low-energetic,
whereas in the $CP$-odd case, the spectrum is much harder,
as discussed previously.

When one imposes a cut on the opening angle of the final-state electron 
momenta w.r.t.\ the beam,
the $CP$-even spectrum becomes harder,
whereas the $CP$-odd one is practically unchanged.
Curves are shown (dashed and dotted) in Fig.~\ref{Fig:eh}, 
corresponding to cuts at opening angles of 5 and 10~degrees. 
(No cut is imposed on the Higgs particle.)
However, even with such a cut, there is a clear distinction 
between the two cases.
We shall return to these distributions in Sec.~8.
%%%%%%%%%%%%%%%%%%%%%%%%%%%%%%%%%%%%%%%%%%%%%%%%%%%%%%%%%%%%%%%%%%%%%%%%
\subsection{Higgs polar-angle distributions}
%%%%%%%%%%%%%%%%%%%%%%%%%%%%%%%%%%%%%%%%%%%%%%%%%%%%%%%%%%%%%%%%%%%%%%%%
Next, we consider the Higgs polar-angle distribution
\begin{equation}
\frac{1}{\sigma^{(h)}}
\frac{\dd\sigma^{(h)}}{\dd\cos\theta_h}
=\frac{1}{\sigma^{(h)}}
\int_{-1}^{1}\dd \cos\theta_1
\int_0^{2\pi}\dd\phi
\int_{m_h}^{E_{h\,{\rm max}}}\dd E_h \,
\frac{\dd^4\sigma^{(h)}}{\dd E_h \,
                     \dd\cos\theta_1\, \dd\cos\theta_h\, \dd\phi} \, .
\end{equation}
The range of integration over $E_h$ is determined by 
$m_h\le E_h\le E+m_h^2/(4E)$.
In Fig.~\ref{Fig:costhh} we show such distributions,
for $\sqrt{s}=500$~GeV, and for two values of the Higgs mass:
$m_h=120$~GeV and 250~GeV.

In the absence of any cut,
there is a clear distinction between the two cases of $CP$,
the $CP$-even distribution being much more peaked along the beam direction.
When cuts are imposed on the opening angles of the final-state
electrons, this difference is reduced, but not seriously.
The dashed and dotted curves in Fig.~\ref{Fig:costhh} show the effects
of imposing a cut on the opening angle of the final-state electrons, 
with respect to the beam, of $5\deg$ and $10\deg$, respectively.
The dependence of these distributions on the Higgs mass, is very weak.

Such polar-angle distributions may therefore be valuable, in particular
if one can get data near the beam directions.

%%%%%%%%%%%%%%%%%%%%%%%%%%%%%%%%%%%%%%%%%%%%%%%%%%%%%%%%%%%%%%%%%%%%%%%%
\subsection{Polarization-dependent correlations}
%%%%%%%%%%%%%%%%%%%%%%%%%%%%%%%%%%%%%%%%%%%%%%%%%%%%%%%%%%%%%%%%%%%%%%%%
The dependence on longitudinal beam polarization enters in the following way
\begin{equation}
\label{Eq:pol-dep}
\dd^4\sigma^{(h)}
=\dd^4\sigma_0^{(h)}
\left[1 +A_1^{(h)}\, P_1P_2 +A_2^{(h)}(P_1+P_2)\right]
\end{equation}
where $|A_1^{(h)}|\le1$, and $|A_2^{(h)}|\le\half(1+A_1^{(h)})$ \cite{OlsOsl}.
These quantities $A_1^{(h)}$ and $A_2^{(h)}$ might be useful in
distinguishing the even and odd case, since the unknown coupling strength
$\eta$ cancels.

The quantity $A_1$ is most easily extracted if both beams have
equal and opposite polarizations.
For the integrated cross section, $A_1$ is shown in 
Fig.~\ref{Fig:sigtot-mh-a1}. 
There is a very strong discrimination between the two $CP$ cases.
For the $CP$-odd case, the cross section is much reduced
if the two beams have large and opposite polarizations,
whereas in the even case, there is only a small reduction.
The large value of $A_1$ in the $CP$-odd case implies that the cross section
is very much reduced if both beams are longitudinally polarized.
This suppression is due to the fact that in the $CP$-odd case,
the two intermediate $Z$s must have orthogonal polarizations.\footnote{We
are grateful to P. Zerwas for this observation.}

In the extraction of $A_1$ from data, there will be a contamination from
the $A_2$ term in (\ref{Eq:pol-dep}) when $P_1+P_2\ne0$.
For the parameters given in Fig.~\ref{Fig:sigtot-mh-a1}, $A_2$
ranges from -16\% to -18\% and from -21\% to -30\% for $CP$ even and odd, 
respectively.
%%%%%%%%%%%%%%%%%%%%%%%%%%%%%%%%%%%%%%%%%%%%%%%%%%%%%%%%%%%%%%%%%%%%%%%%
\section{Final-state electron-electron correlations}
\setcounter{equation}{0}
\label{section:ee-corr}
%%%%%%%%%%%%%%%%%%%%%%%%%%%%%%%%%%%%%%%%%%%%%%%%%%%%%%%%%%%%%%%%%%%%%%%%
In the electron-electron mode, the angular distributions are more
complicated than in the positron-electron mode, 
due to the fact that the propagators will depend
on the angles of interest, through $t_j$ and $u_j$.
%%%%%%%%%%%%%%%%%%%%%%%%%%%%%%%%%%%%%%%%%%%%%%%%%%%%%%%%%%%%%%%%%%%%%%%%
\subsection{Azimuthal correlations}
%%%%%%%%%%%%%%%%%%%%%%%%%%%%%%%%%%%%%%%%%%%%%%%%%%%%%%%%%%%%%%%%%%%%%%%%

We first consider distributions in the azimuthal angle $\phi$
defined in (\ref{Eq:phi}). These are
obtained by integrating the differential cross section, 
Eq.~(\ref{EQ:1-dsigma}),
over the energy difference, given by $\varepsilon$,
up to $\varepsilonmax$,
as well as over the polar angles, $\theta_j$, 
for $0\le|\cos\theta_j|\le\cos\theta_c$:
\begin{eqnarray}
&&\frac{2\pi}{\sigma^{(h)}[\cos\theta_c]}
\frac{\dd\sigma^{(h)}[\cos\theta_c]}{\dd\phi} \nonumber \\
&&=\frac{2\pi}{\sigma^{(h)}[\cos\theta_c]}
\int_0^{\varepsilonmax}\dd\varepsilon
\int_{-\cos\theta_c}^{\cos\theta_c}\dd\cos\theta_1 
\int_{-\cos\theta_c}^{\cos\theta_c}\dd\cos\theta_2
\frac{\dd^4\sigma^{(h)}}{\dd\varepsilon\,
                     \dd\cos\theta_1\, \dd\cos\theta_2\, \dd\phi},
\end{eqnarray}
with
\begin{equation}
\sigma^{(h)}[\cos\theta_c]
=\int_0^{\varepsilonmax}\dd\varepsilon
\int_{-\cos\theta_c}^{\cos\theta_c}\dd\cos\theta_1 
\int_{-\cos\theta_c}^{\cos\theta_c}\dd\cos\theta_2
\int_0^{2\pi}\dd\phi\,
\frac{\dd^4\sigma^{(h)}}{\dd\varepsilon\,
                     \dd\cos\theta_1\, \dd\cos\theta_2\, \dd\phi}.
\end{equation}

In this case, there is no particular need to use the events where
the final-state electrons are close to the beam direction,
so we impose a stronger cut, $\cos\theta_c=0.9$.
(It may even be difficult to determine the azimuthal angles
for electrons which are close to the beam direction.)
Results are shown in Fig.~\ref{Fig:phi-sig}, for the unpolarized case.
We consider two c.m.~energies, and two Higgs masses.
The distributions generally favor the region around $\phi\sim\pi$,
i.e., when the two final-state electrons have non-vanishing
and opposite transverse momenta (w.r.t.\ the beam), as opposed
to $\phi\sim0$, when they are more parallel.
This broad feature is purely kinematic, more energy is available
to create a Higgs particle if the two virtual $Z$s have opposite 
transverse momenta.
On top of this broad feature, there is in the $CP$-odd case
a dip around $\phi=\pi$, if the Higgs momentum is sufficiently high
(i.e., at low mass).

If the two beams have opposite polarizations, the difference
between the two cases can be quite spectacular, as is illustrated in 
Fig.~\ref{Fig:phi-sig-500-120-pm} for $CP=1$ and $CP=-1$.
%%%%%%%%%%%%%%%%%%%%%%%%%%%%%%%%%%%%%%%%%%%%%%%%%%%%%%%%%%%%%%%%%%%%%%%%
\subsection{Polar-angle correlations}
%%%%%%%%%%%%%%%%%%%%%%%%%%%%%%%%%%%%%%%%%%%%%%%%%%%%%%%%%%%%%%%%%%%%%%%%

Next we consider distributions in the polar angles of the electrons,
\begin{equation}
\frac{1}{\sigma^{(h)}[\cos\theta_c]}
\frac{\dd^2\sigma^{(h)}}{\dd\cos\theta_1 \dd\cos\theta_2}
=\frac{1}{\sigma^{(h)}[\cos\theta_c]}
\int_0^{\varepsilonmax}\dd\varepsilon\int_0^{2\pi}\dd\phi\,
\frac{\dd^4\sigma^{(h)}}{\dd\varepsilon\,
                     \dd\cos\theta_1\, \dd\cos\theta_2\, \dd\phi}
\end{equation}
Such distributions are shown in Figs.~\ref{Fig:cos-cos}, 
for $CP$ even and odd, and for the case of no polarization.
There is a rather strong difference between the two cases,
the cross section being much more peaked for electrons
emitted close to the forward direction in the $CP$ even case.
To produce an odd parity state, angular momentum
has to be transferred, and the electrons must therefore undergo
a more violent scattering.

A less differential distribution can be obtained as follows.
Let 
\begin{equation}
\cos\Theta=\half(\cos\theta_1-\cos\theta_2),
\end{equation}
or
\begin{equation}
\cos\theta_1=\cos\Theta +\half w, \qquad 
\cos\theta_2=-\cos\Theta +\half w,
\end{equation}
and consider
\begin{equation}
\frac{1}{\sigma^{(h)}[\cos\theta_c]}
\frac{\dd\sigma^{(h)}}{\dd\cos\Theta}
=\frac{1}{\sigma^{(h)}[\cos\theta_c]}
\int_0^{\varepsilonmax}\dd\varepsilon\int_0^{2\pi}\dd\phi\,
\int_{-\wmax}^{\wmax}\dd w\,
\frac{\dd^4\sigma^{(h)}}{\dd\varepsilon\,
                     \dd\cos\theta_1\, \dd\cos\theta_2\, \dd\phi}
\end{equation}
with 
\begin{equation}
\wmax=
\begin{cases}
2(\cos\theta_c+\cos\Theta) & \text{if $\cos\Theta<0$},\\
2(\cos\theta_c-\cos\Theta) & \text{if $\cos\Theta>0$}.
\end{cases}
\end{equation}
We show in Fig.~\ref{Fig:cos12-pol} such distributions, 
for the even and odd cases.
For $CP=1$, the cross section is much more peaked towards
the ``forward" direction, $\cos\Theta=\pm1$, consistent
with Fig.~\ref{Fig:cos-cos}.
%%%%%%%%%%%%%%%%%%%%%%%%%%%%%%%%%%%%%%%%%%%%%%%%%%%%%%%%%%%%%%%%%%%%%%%%
\subsection{Energy correlations}
%%%%%%%%%%%%%%%%%%%%%%%%%%%%%%%%%%%%%%%%%%%%%%%%%%%%%%%%%%%%%%%%%%%%%%%%

Finally, we consider the distribution in relative electron-energy 
difference.
Introducing the scaled energy difference as 
$x=\varepsilon/\varepsilonmax$ [see Eqs.~(\ref{Eq:varepsilon})
and(\ref{Eq:varepsilonmax})], we will consider 
\begin{equation}
\frac{1}{\sigma^{(h)}[\cos\theta_c]}
\frac{{\rm d}\sigma^{(h)}[\cos\theta_c]}{{\rm d} x}.
\end{equation}
Such distributions are shown in Fig.~\ref{Fig:eps-sig-pol00}.
For the $CP$-odd case, this distribution is ``harder",
it falls off less rapidly for large energy differences $x$.

In electron-positron annihilation, with Higgs production
via the Bjorken process, analogous distributions also exhibit
a considerable sensitivity to whether
the Higgs particle is even or odd under $CP$ \cite{SkjOsl95}.
%%%%%%%%%%%%%%%%%%%%%%%%%%%%%%%%%%%%%%%%%%%%%%%%%%%%%%%%%%%%%%%%%%%%%%
\section{$CP$ violation}
\setcounter{equation}{0}
%%%%%%%%%%%%%%%%%%%%%%%%%%%%%%%%%%%%%%%%%%%%%%%%%%%%%%%%%%%%%%%%%%%%%%
As discussed in the Introduction, there could also be $CP$-violation
in the Higgs sector, in which case the Higgs particles would not
be eigenstates of $CP$.

While the presence of both even ($X^{(H)}$ and $Z^{(H)}$) and
odd ($X^{(A)}$ and $Z^{(A)}$) terms in the cross section
reflect parity violation,
only the terms $\tilde X_t$ and $\tilde X_u$ explicitly violate
parity. For these to be observed, one has to assign a value to 
$\vecp_1'\times\vecp_2'$ [cf.\ Eq.~(\ref{Eq:H-cp-viol})],
i.e., one needs to distinguish the final-state electrons.

We show in Fig.~\ref{Fig:phi-sig-cp} azimuthal distributions of the
kind shown in Fig.~\ref{Fig:phi-sig}, allowing for $CP$ violation.
Since these involve a symmetrical integration over both hemispheres,
$-\cos\theta_c\le\cos\theta_{1,2}\le\cos\theta_c$,
the parity-violating terms $\tilde X_t$ and $\tilde X_u$
[cf.\ Eq.~(\ref{Eq:X-tu-tilde})] cancel.
However, the parity violation leads to a superposition of the
two cases, $CP=+1$ and $CP=-1$.
Such distributions may suffice to provide evidence of parity violation.

One way to access the parity-violating terms $\tilde X_t$ and $\tilde X_u$,
is to introduce the weight factor $\cos\theta_1$ to distinguish
the two hemispheres. Thus, we consider (the two electrons are
here distinguished by $E_1'>E_2'$) the asymmetry
\begin{equation}
\label{Eq:CP-A}
A=\frac{2\pi}{\sigma^{(h)}[\cos\theta_c]}
\int_0^{\varepsilonmax}\dd\varepsilon
\int_{-\cos\theta_c}^{\cos\theta_c}\dd\cos\theta_1 
\int_{-\cos\theta_c}^{\cos\theta_c}\dd\cos\theta_2
\frac{\dd^4\sigma^{(h)}\,\cos\theta_1}{\dd\varepsilon\,
                     \dd\cos\theta_1\, \dd\cos\theta_2\, \dd\phi}.
\end{equation}
This quantity is shown in Fig.~\ref{Fig:phi-sig-cp-as} for the same
parameters and cuts as were used in Fig.~\ref{Fig:phi-sig-cp}.
To lowest order, the effect is linear in $\eta$.
Thus, given enough data, the effect can be sizable.
Other ways to search for $CP$ violation in the $ZZ$-Higgs coupling
are discussed in \cite{e-p,SkjOsl95,NelCha}
%%%%%%%%%%%%%%%%%%%%%%%%%%%%%%%%%%%%%%%%%%%%%%%%%%%%%%%%%%%%%%%%%%%%%%
\section{Charged Higgs Production}
\setcounter{equation}{0}
%%%%%%%%%%%%%%%%%%%%%%%%%%%%%%%%%%%%%%%%%%%%%%%%%%%%%%%%%%%%%%%%%%%%%%
If the produced Higgs is charged, there will be one or two
final-state neutrinos. These cannot be detected, so distributions 
of the kind discussed in section \ref{section:ee-corr} are not available.
One may instead consider distributions of the charged Higgs
particles themselves.
%%%%%%%%%%%%%%%%%%%%%%%%%%%%%%%%%%%%%%%%%%%%%%%%%%%%%%%%%%%%%%%%%%%%%%%%
\subsection{Singly-charged Higgs production}
%%%%%%%%%%%%%%%%%%%%%%%%%%%%%%%%%%%%%%%%%%%%%%%%%%%%%%%%%%%%%%%%%%%%%%%%
Singly-charged (negative) Higgs particles, which are expected in 
certain models \cite{HHG}, 
can be produced in $e^-e^-$ collisions through the 
exchange of one $Z$ and one $W^-$ boson. The cross section would be 
given by formulas analogous to those presented in Sec.~3, where the
numerical coefficients involving the polarizations $P_i$ and the relative
strength of the axial coupling, $\sin2\chi$, would be replaced
as follows for the $t$-channel terms (with accompanying changes in
the propagator masses):
\begin{eqnarray}
(1-P_1\sin2\chi)(1-P_2\sin2\chi)
&\to&(1-P_1\sin2\chi)(1-P_2), \nonumber \\
(\sin2\chi-P_1)(\sin2\chi-P_2)
&\to&(\sin2\chi-P_1)(1-P_2),
\end{eqnarray}
and similarly for the $u$-channel terms,
with $P_1$ and $P_2$ interchanged.
The interference terms would have the coefficient
substitution
\begin{equation}
(1+P_1P_2)(1+\sin^2{2\chi})-2(P_1+P_2)\sin2\chi\to
(1-P_1)(1-P_2)(1+\sin2\chi),
\nonumber
\end{equation}
where $\sin2\chi\simeq 0.1393$ refers to the $eeZ$ coupling.

These substitutions would only change quantitative aspects of the
cross sections. Thus, we expect all qualitative features discussed 
in Sec.~4 to remain valid.

%%%%%%%%%%%%%%%%%%%%%%%%%%%%%%%%%%%%%%%%%%%%%%%%%%%%%%%%%%%%%%%%%%%%%%%%
\subsection{Doubly-charged Higgs production}
%%%%%%%%%%%%%%%%%%%%%%%%%%%%%%%%%%%%%%%%%%%%%%%%%%%%%%%%%%%%%%%%%%%%%%%%
Doubly-charged Higgs particles, $h^{--}$, which are expected in the
left--right-symmetric \cite{Mohapatra} and other models \cite{double-h}, 
can be produced in electron-electron collisions, not only in the
$s$-channel, but also via $WW$ exchange.
This mechanism does not require lepton-number violation,
but the $WWh^{--}$ coupling is absent in certain models \cite{Gunion}.
Apart from an over-all, model-dependent constant, the cross section
would be given by the formulas of Sec.~3, with $\sin2\chi=1$.
Distributions of the kinds given in Figs.~\ref{Fig:eh} 
and \ref{Fig:sigtot-mh-a1} would readily
reveal whether such a particle was even or odd under $CP$.

%%%%%%%%%%%%%%%%%%%%%%%%%%%%%%%%%%%%%%%%%%%%%%%%%%%%%%%%%%%%%%%%%%%%%%
\section{Statistical considerations}
\setcounter{equation}{0}
%%%%%%%%%%%%%%%%%%%%%%%%%%%%%%%%%%%%%%%%%%%%%%%%%%%%%%%%%%%%%%%%%%%%%%

It is of interest to estimate how many events are
needed to determine the $CP$ from distributions 
of the kinds presented here.
One of the most promising ones appears to be the Higgs energy
distribution, shown in Fig.~\ref{Fig:eh}. We will assume that
$CP$ is conserved, so that the problem can be formulated in terms
of statistical hypothesis testing as $H_0$: $CP=1$ and
$H_1$: $CP=-1$. 
Information would be gained if $H_0$ were rejected.

We denote the $CP=+1$ distribution by $f(x)$ and
the $CP=-1$ distribution by $g(x)$.
The problem is well suited for the Neyman-Pearson test \cite{NePe},
and following this approach, we will reject $H_0$ if,
for $n$ events, the likelihood ratio
\begin{equation}
L(x_1,x_2,\dots,x_n)=\frac{g(x_1)g(x_2)\cdots g(x_n)}
{f(x_1)f(x_2)\cdots f(x_n)}\geq k,
\label{nepe}
\end{equation}
where $x_i$ denote observed values of $E_h$, and
$k$ is a critical constant to be determined.
The constant $k$ determines the level of ``significance'', $\alpha$,
of the test, i.e., the probability that we reject a {\it correct} hypothesis.
An estimate for $k$ can be obtained by Monte Carlo simulations.
By drawing $s$ samples of $n$ $x$-values from the $f(x)$ distribution,
the ratios $L_1$, $L_2$,\ldots $L_s$ can be calculated
by applying Eq.~(\ref{nepe}).
The empirical $(1-\alpha)\cdot100\%$ percentile in the simulated $L$
distribution can be used as an estimate of $k$.

The ``power'', $1-\beta$ (the probability of rejecting $H_0$ when $H_0$
is false), of this resulting test can also be estimated 
by Monte Carlo simulations.
Samples should then be drawn from the $g$ distribution,
and the proportion of samples that are rejected in the test estimates
$1-\beta$. Results are given in table~1
for $E_{\rm cm}=500$~GeV and three values of the Higgs mass,
$m_h=100$, 120 and 150~GeV.
For each mass value, two cases are considered:
(i) No cut on the final-state electron momenta,
(ii) the electron momenta have to satisfy $\theta\ge10\deg$.
(With a cut at $5\deg$, these probabilities are practically
the same as without any cut.)
We see that at a mass of 120~GeV,
already 10 events suffice to reveal a $CP=-1$ distribution,
at the level of 95--97\%, with a risk of falsely rejecting 
the correct hypothesis, $\alpha$, of only 5\%.
The discrimination is easier if one can get data near the beam direction, 
and if the Higgs particle is light, as is also seen from Fig.~\ref{Fig:eh}.
\medskip

\begin{table}
\begin{center}
%%%%%%%%%%%%%%%%%%%%%%%%%%%%%%%%%%%%%%%%%%%%%%%%%%%%%%%%%%%%%%%%%%%%%%%%
\begin{tabular}{||c|r|r|r|c|r|c|r|c||}  \hline
 &  &   & \multicolumn{2}{c|}{100~GeV} & \multicolumn{2}{c|}{120~GeV} &
          \multicolumn{2}{c||}{150~GeV} \\
\cline{4-9}
 &  $n$      &    $\alpha$ [\%]   
& $k$ & $1-\beta$ [\%] & $k$ & $1-\beta$ [\%] & $k$ & $1-\beta$ [\%] \\ \hline
\multirow{2}{20mm}{No cut} 
& 5  &   5.0  &   3.0   & 81  & 3.2    &  80 &   3.5  &  76   \\
&10  &   5.0  &   0.75  & 97  & 0.94   &  97 &   1.2  &  95   \\ \hline
\multirow{2}{20mm}{Cut at $10\deg$} 	
& 5  &   5.0  &   3.2   & 80  & 3.4    &  76 &   3.6  &  72   \\
&10  &   5.0  &   1.0   & 96  & 1.2    &  95 &   1.5  &  93   \\ \hline\hline
\end{tabular}
%%%%%%%%%%%%%%%%%%%%%%%%%%%%%%%%%%%%%%%%%%%%%%%%%%%%%%%%%%%%%%%%%%%%%%%%
\end{center}
\caption{Recognition probabilities.
Here, $n$ is the number of events, $\alpha$ is the level of significance 
of the test and $1-\beta$ the probability that
one can recognize a $CP=-1$ distribution in Higgs energy data,
for $E_{\rm cm}=500$~GeV and at three Higgs masses, $m_h=100$,
120 and 150~GeV.
See the text for further details.}
\end{table}
\medskip

%%%%%%%%%%%%%%%%%%%%%%%%%%%%%%%%%%%%%%%%%%%%%%%%%%%%%%%%%%%%%%%%%%%%%%
\section{Concluding remarks}
\setcounter{equation}{0}
%%%%%%%%%%%%%%%%%%%%%%%%%%%%%%%%%%%%%%%%%%%%%%%%%%%%%%%%%%%%%%%%%%%%%%
We have studied the production of generic Higgs particles in $e^-e^-$
collisions, focusing on distributions which might be useful in
distinguishing a $CP$-even from a $CP$-odd particle.
Longitudinal beam polarization effects are taken into account.

We have not discussed backgrounds. These would depend on how
the Higgs boson is detected.
A light Higgs would dominantly decay to $b$ quarks, and the
background would not be severe, mostly from single $Z$ and $W$ production
and the two-photon process \cite{Barger}.
A heavier Higgs would decay to $W$ and $Z$ bosons, and the background
would be a problem \cite{Hikasa,Cuypers}.

In the $CP$-even case, the Higgs particle tends to be softer, 
and events are more aligned with the beam direction than in 
the $CP$-odd case.
In fact, the Higgs energy distribution may be one of the better
observables for discriminating the two cases.

Furthermore, the dependence on the {\it product} of the two
beam polarizations is much larger in the $CP$-odd case.
This dependence, which is represented by an observable $A_1$,
becomes a better ``discriminator'' for increasing Higgs masses,
when the Higgs momentum decreases, and other methods may tend 
to become less efficient.

If the two final-state electrons are observed, a certain azimuthal
distribution, as well as the electron polar-angle distributions,
will also be useful for discriminating the two cases.

Finally, we suggest ways to search for possible parity-violating effects 
in the $ZZ$-Higgs coupling.
%%%%%%%%%%%%%%%%%%%%%%%%%%%%%%%%%%%%%%%%%%%%%%%%%%%%%%%%%%%%%%%%%%%%%%

\vspace{10mm}

{\bf Acknowledgments.}
It is a pleasure to thank C. Irgens, P. Minkowski, C. Newton and T. T. Wu 
for most valuable discussions.
This research has been supported by the Research Council of Norway.
JZZ would like to thank the Department of Physics, University
of Bergen, for warm hospitality.
His work has been supported by the National Natural Science Foundation 
of China under Grant No.\ 19674014, and the Shanghai Education
Development Foundation.

\clearpage
\appendix
%%%%%%%%%%%%%%%%%%%%%%%%%%%%%%%%%%%%%%%%%%%%%%%%%%%%%%%%%%%%%%%%%%%%%%
\section*{Appendix A. The $e^+e^-$ cross section}
\renewcommand{\thesection}{A}
\setcounter{equation}{0}
%%%%%%%%%%%%%%%%%%%%%%%%%%%%%%%%%%%%%%%%%%%%%%%%%%%%%%%%%%%%%%%%%%%%%%

For the positron-electron case, there is, in addition to the familiar
Bjorken diagram, also a $t$-channel diagram.
The cross section can be expressed as
\begin{eqnarray}
\frac{\dd^4\sigma^{(h)}}{\dd\varepsilon\,
                     \dd\cos\theta_1\, \dd\cos\theta_2\, \dd\phi}
&=&C^{(h)}\bigl\{[F(t_1,t_2)]^2\, X^{(h)}
+ [F(s_1,s_2)]^2\, \tilde X^{(h)} \nonumber \\
& & \phantom{mm} + 2 \Re [F^*(s_1,s_2) F(t_1,t_2)] Z^{(h)}\bigr\},
\end{eqnarray}
with $F(t_1,t_2)$ defined by Eq.~(\ref{eq-2-F}).

The amplitude for the $t$-channel diagram is related to the corresponding
one for the electron-electron case by the spinor substitutions
$\bar v(p_1)\to \bar u(p_1')$, and $v(p_1')\to u(p_1)$,
which amount to $(s_1,s_2)\leftrightarrow(-u_2,-u_1)$.
Also, the convention for the positron polarization is different, 
such that $P_1\to -P_1$.

Furthermore, the ($s$-channel) Bjorken diagram is related to the
$t$-channel diagram in a way similar to what is the case 
for the electron-electron diagrams. Thus, the unpolarized cross
section for the Bjorken diagram is related to that of the $t$-channel
diagram by $(s_1,s_2)\leftrightarrow(-t_1,-t_2)$.
However, the positron polarization for the Bjorken diagram would 
correspond to a final-state polarization
in the $t$-channel diagram (which we sum over).
Thus, the polarization-dependent parts of these cross sections
are not related in this simple way.

For the $CP$-even case, we find
\begin{eqnarray}
X^{(H)}&=&
2\bigl[(1+P_1\sin2\chi)(1-P_2\sin2\chi)(s_1 s_2 +u_1 u_2) \nonumber \\
& & \phantom{2\bigl[}
-(\sin2\chi+P_1)(\sin2\chi-P_2)(s_1 s_2 -u_1 u_2)\bigr] \nonumber \\
&=&2[(1+P_1P_2)(1-\sin^{2}2\chi)s_1s_2
+(1-P_1P_2)(1+\sin^{2}2\chi)u_1u_2 \nonumber \\
& & \phantom{2\bigl[}
+2(P_1-P_2)\sin2\chi u_1u_2], \\
\tilde X^{(H)}&=&
2\{(1-P_1P_2)[t_1t_2+u_1u_2-\sin^{2}2\chi(t_1t_2-u_1u_2)] \nonumber \\
& & \phantom{2\{}
+2(P_1-P_2)\sin2\chi u_1u_2\}, \\
Z^{(H)} &=&
%2\bigl[ (1+P_1\sin2\chi) (1-P_2\sin2\chi)
%       + (\sin2\chi+P_1) (\sin2\chi-P_2) \bigr]u_1 u_2,
2[(1-P_1P_2)(1+\sin^{2}2\chi)+2(P_1-P_2)\sin2\chi] u_1u_2.
\end{eqnarray}

For the $CP$-odd case, we find
\begin{eqnarray}
X^{(A)}&=&
\frac{t_1t_2}{2}\left[(s_1+s_2)^2+(u_1+u_2)^2\right]
-(s_1s_2-u_1u_2)^2-(t_1t_2)^2 \nonumber \\
& & -\sin^22\chi\;
\frac{t_1t_2}{2}\left[(s_1-s_2)^2-(u_1-u_2)^2\right] \nonumber \\
& & +(P_1-P_2)\sin2\chi
\Bigl[ t_1t_2(u_1^2+u_2^2)-(s_1s_2)^2-(t_1t_2)^2-(u_1u_2)^2 \nonumber \\
& & \phantom{+(P_1-P_2)\sin2\chi}
      +2s_1s_2(t_1t_2+u_1u_2) \Bigr] \nonumber \\
& & +P_1P_2\biggl\{ \frac{t_1t_2}{2}
[(s_1-s_2)^2-(u_1-u_2)^2] \nonumber \\
& & +\sin^22\chi\,
\biggl(-\frac{t_1t_2}{2}\left[(s_1+s_2)^2+(u_1+u_2)^2\right]
+(s_1s_2-u_1u_2)^2+(t_1t_2)^2 \biggr)\biggr\}, \nonumber \\
\end{eqnarray}
\begin{eqnarray}
\tilde X^{(A)}&=&
\frac{s_1s_2}{2}\left[(t_1+t_2)^2+(u_1+u_2)^2\right]
-(t_1t_2-u_1u_2)^2-(s_1s_2)^2 \nonumber \\
& & -\sin^22\chi\frac{s_1s_2}{2}[(t_1-t_2)^2 -(u_1-u_2)^2] \nonumber \\
& & +(P_1-P_2)\sin2\chi
\Bigl[s_1s_2(u_1^2+u_2^2)-(s_1s_2)^2-(t_1t_2)^2-(u_1u_2)^2 \nonumber \\
& & \phantom{+(P_1-P_2)\sin2\chi}
+2t_1t_2(s_1s_2+u_1u_2)\Bigr] \nonumber \\
& & +P_1P_2\biggl\{ -\frac{s_1s_2}{2}
[(t_1+t_2)^2+(u_1+u_2)^2] 
+(t_1t_2-u_1u_2)^2+(s_1s_2)^2 \nonumber \\
& & \phantom{+P_1P_2\biggl\{}
+\sin^22\chi\frac{s_1s_2}{2}[(t_1-t_2)^2-(u_1-u_2)^2]\biggr\}, \\
Z^{(A)}&=&
\frac{1}{4}\bigl[(1-P_1P_2)(1+\sin^22\chi) +2(P_1-P_2)\sin2\chi\bigr] 
\nonumber \\
& & \times
\bigl[2(s_1s_2-t_1t_2)^2 +u_1u_2(u_1^2+u_2^2)
     -(s_1s_2+t_1t_2)(u_1+u_2)^2 \bigr].
\end{eqnarray}

%%%%%%%%%%%%%%%%%%%%%%%%%%%%%%%%%%%%%%%%%%%%%%%%%%%%%%%%%%%%%%%%%%%%%%%%

\newpage
%%%%%%%%%%%%%%%%%%%%%%%%%%%%%%%%%%%%%%%%%%%%%%%%%%%%%%%%%%%%%%%%%%%%
\begin{figure}[htb]
\refstepcounter{figure}
\label{Fig:feynman}
\addtocounter{figure}{-1}
\phantom{AAA}
\begin{center}
\setlength{\unitlength}{1cm}
\begin{picture}(18,20)
\put(0.0,0.0)
{\mbox{\epsfysize=20cm\epsffile{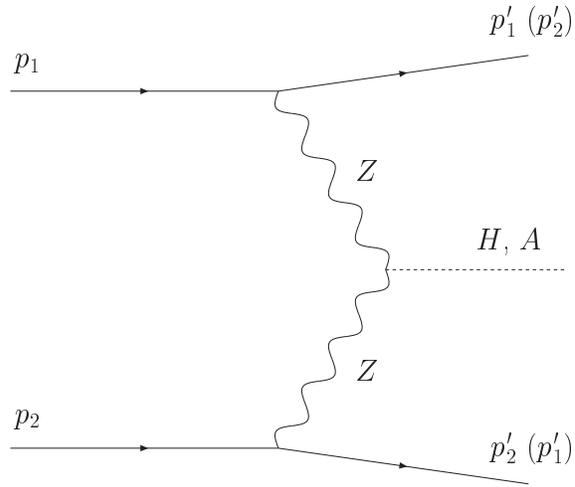}}}
\end{picture}
\end{center}
\vspace*{-8.0cm}
\caption{Feynman diagram for the class of processes considered.
(There is also a crossed diagram.)}
\end{figure}
%%%%%%%%%%%%%%%%%%%%%%%%%%%%%%%%%%%%%%%%%%%%%%%%%%%%%%%%%%%%%%%%%%%%
\newpage

%%%%%%%%%%%%%%%%%%%%%%%%%%%%%%%%%%%%%%%%%%%%%%%%%%%%%%%%%%%%%%%%%%%%
\begin{figure}[htb]
\refstepcounter{figure}
\label{Fig:sigtot-mh}
\addtocounter{figure}{-1}
\phantom{AAA}
\begin{center}
\setlength{\unitlength}{1cm}
\begin{picture}(16,20)
\put(3.0,5.0)
{\mbox{\epsfysize=9cm\epsffile{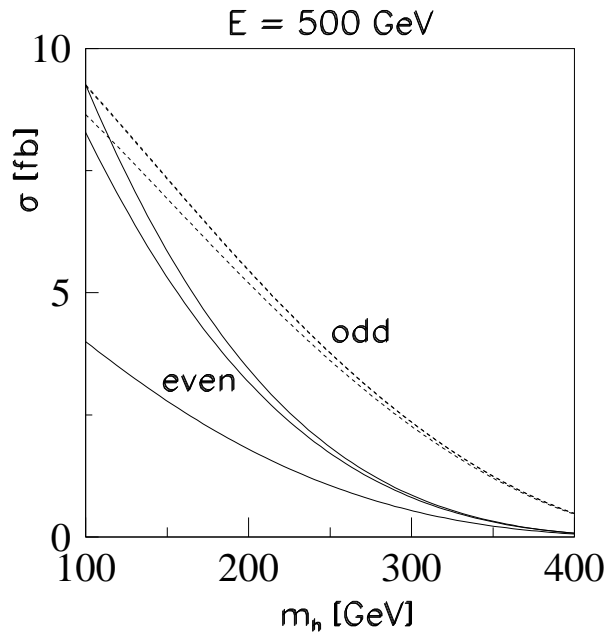}}}
\end{picture}
\end{center}
\vspace*{-3.0cm}
\caption{Cross sections for Higgs production in electron-electron
collisions at $E_{\rm c.m.}=500$~GeV, for a range of Higgs masses.
Standard Model (denoted ``even'') and $CP$-odd results are shown.
For each case, the upper curve corresponds to no cut,
whereas the middle and lower ones are obtained with angular cuts
at $5\deg$ and $15\deg$, respectively.
(In the odd case, the curve for $5\deg$ cannot be distinguished from
the one for no cut.)
The cross sections for the odd case are normalized such that 
for no cuts, they coincide at $m_h=100$~GeV, yielding $\eta=0.884$ 
[see Eq.~(\ref{Eq:coupling})].}
\end{figure}
%%%%%%%%%%%%%%%%%%%%%%%%%%%%%%%%%%%%%%%%%%%%%%%%%%%%%%%%%%%%%%%%%%%%

%%%%%%%%%%%%%%%%%%%%%%%%%%%%%%%%%%%%%%%%%%%%%%%%%%%%%%%%%%%%%%%%%%%%
\begin{figure}[htb]
\refstepcounter{figure}
\label{Fig:sigtot-e}
\addtocounter{figure}{-1}
\phantom{AAA}
\begin{center}
\setlength{\unitlength}{1cm}
\begin{picture}(16,20)
\put(3.0,5.0)
{\mbox{\epsfysize=9cm\epsffile{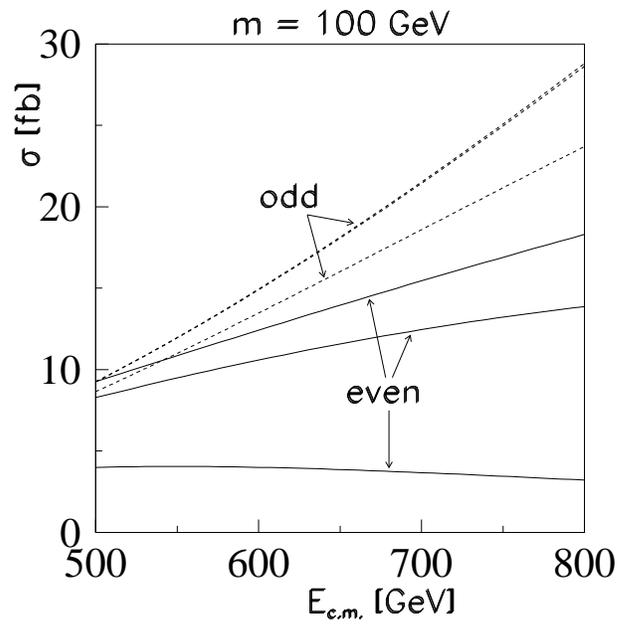}}}
\end{picture}
\end{center}
\vspace*{-3.0cm}
\caption{Cross sections for Higgs production in electron-electron
collisions for a Higgs mass $m_h=100$~GeV, for a range of 
energies, $E_{\rm c.m.}$.
Standard Model (denoted ``even'') and $CP$-odd results are shown.
For each case, the upper curve corresponds to no cut,
whereas the lower ones are obtained with the same angular cuts
as in Fig.~\ref{Fig:sigtot-mh}.
The cross sections for the odd case are normalized like in Fig.~2.}
\end{figure}
%%%%%%%%%%%%%%%%%%%%%%%%%%%%%%%%%%%%%%%%%%%%%%%%%%%%%%%%%%%%%%%%%%%%

%%%%%%%%%%%%%%%%%%%%%%%%%%%%%%%%%%%%%%%%%%%%%%%%%%%%%%%%%%%%%%%%%%%%
\begin{figure}[htb]
\refstepcounter{figure}
\label{Fig:eh}
\addtocounter{figure}{-1}
\phantom{AAA}
\begin{center}
\setlength{\unitlength}{1cm}
\begin{picture}(16,20)
\put(3.0,12.0)
{\mbox{\epsfysize=8cm\epsffile{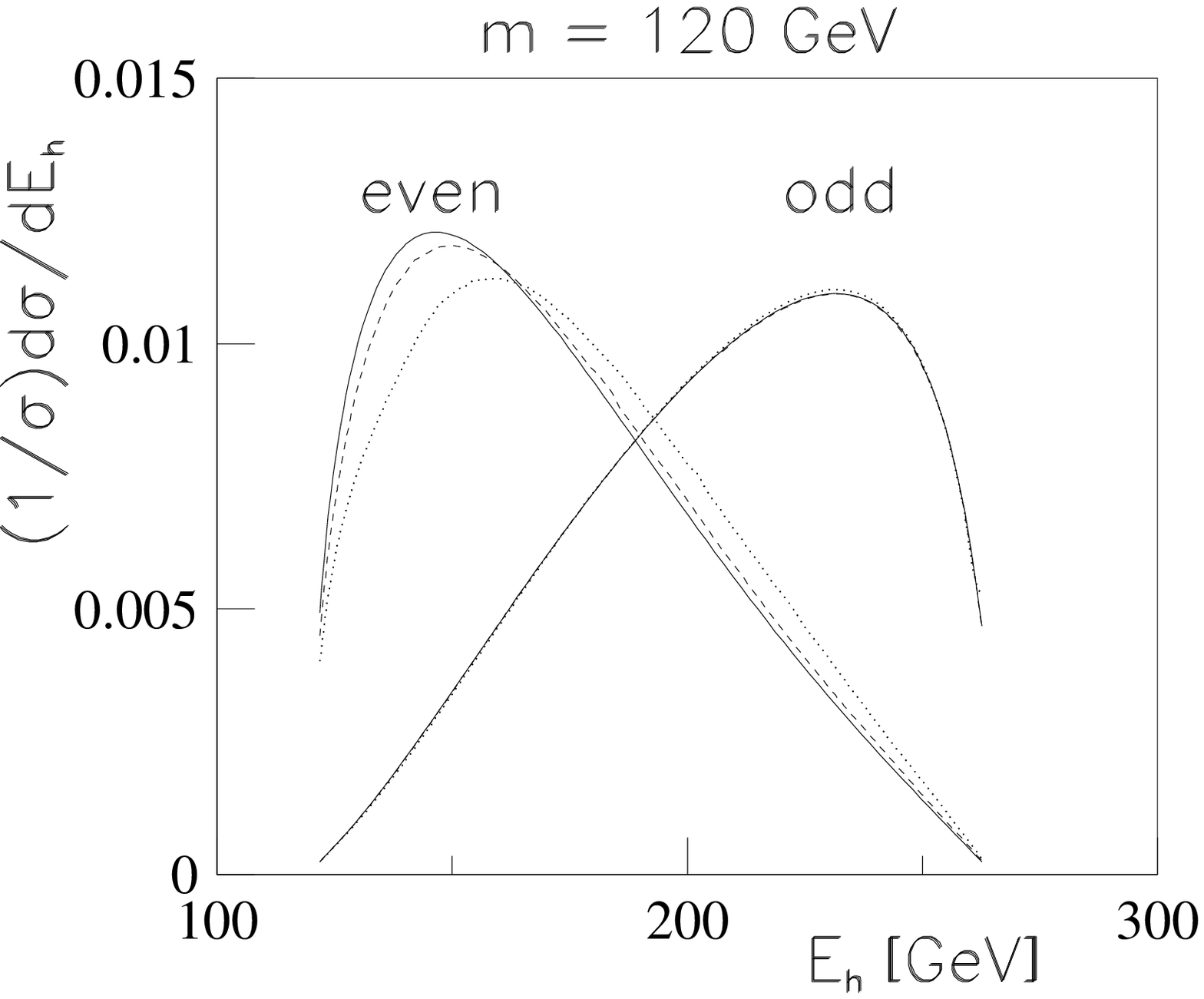}}}
\put(3.0,3.0)
{\mbox{\epsfysize=8cm\epsffile{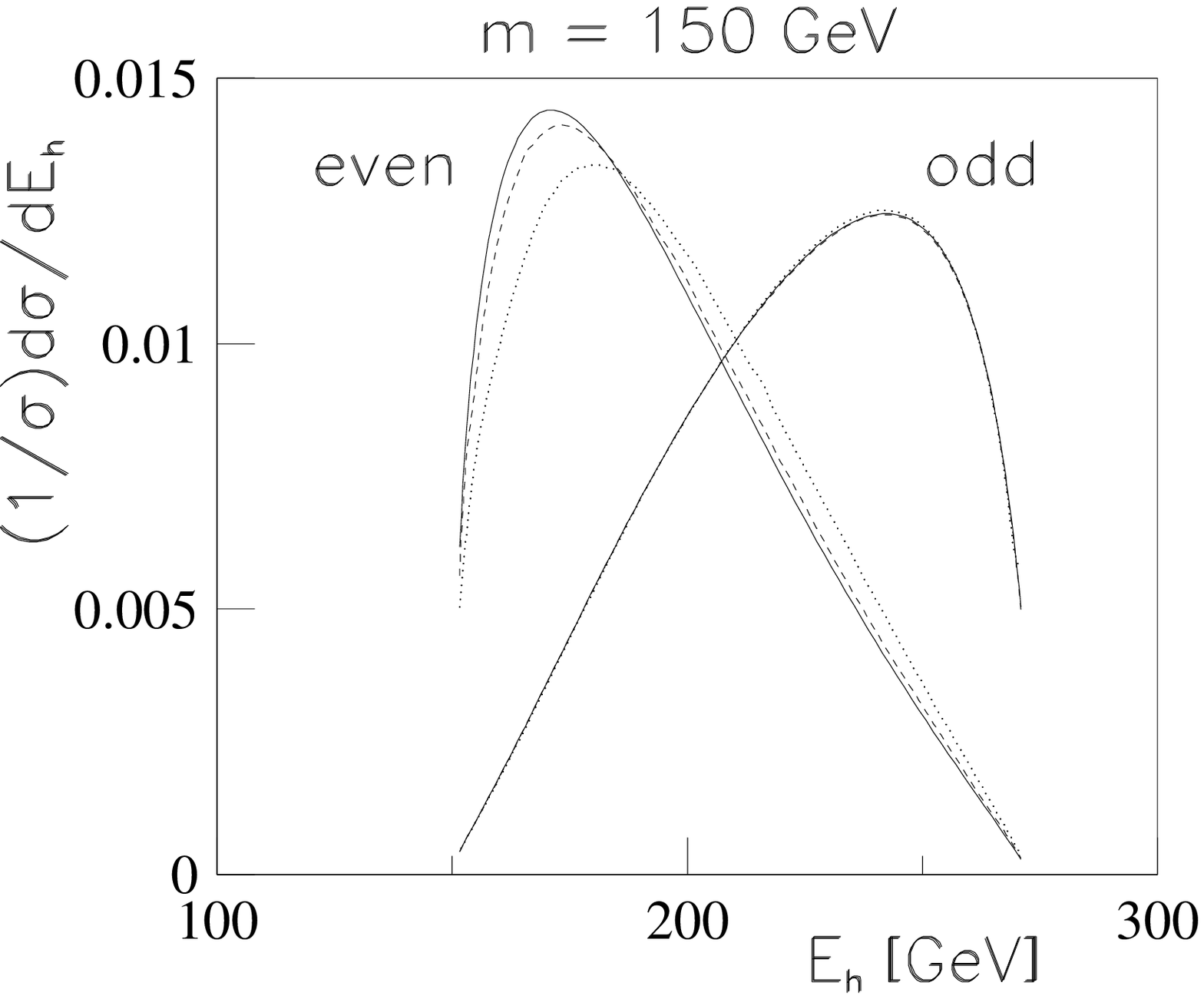}}}
\end{picture}
\end{center}
\vspace*{-2.0cm}
\caption{Higgs energy spectra for the case $E_{\rm c.m.}=500$~GeV,
and for Higgs masses $m_h=120$~GeV and 150~GeV.
The solid curves give the distributions in the absence of any cut.
The dashed and dotted curves show the corresponding distributions when 
cuts at $5\deg$ and $10\deg$ are imposed on the electron momenta.}
\end{figure}
%%%%%%%%%%%%%%%%%%%%%%%%%%%%%%%%%%%%%%%%%%%%%%%%%%%%%%%%%%%%%%%%%%%%

%%%%%%%%%%%%%%%%%%%%%%%%%%%%%%%%%%%%%%%%%%%%%%%%%%%%%%%%%%%%%%%%%%%%
\begin{figure}[htb]
\refstepcounter{figure}
\label{Fig:costhh}
\addtocounter{figure}{-1}
\phantom{AAA}
\begin{center}
\setlength{\unitlength}{1cm}
\begin{picture}(16,20)
\put(3.0,12.0)
{\mbox{\epsfysize=8cm\epsffile{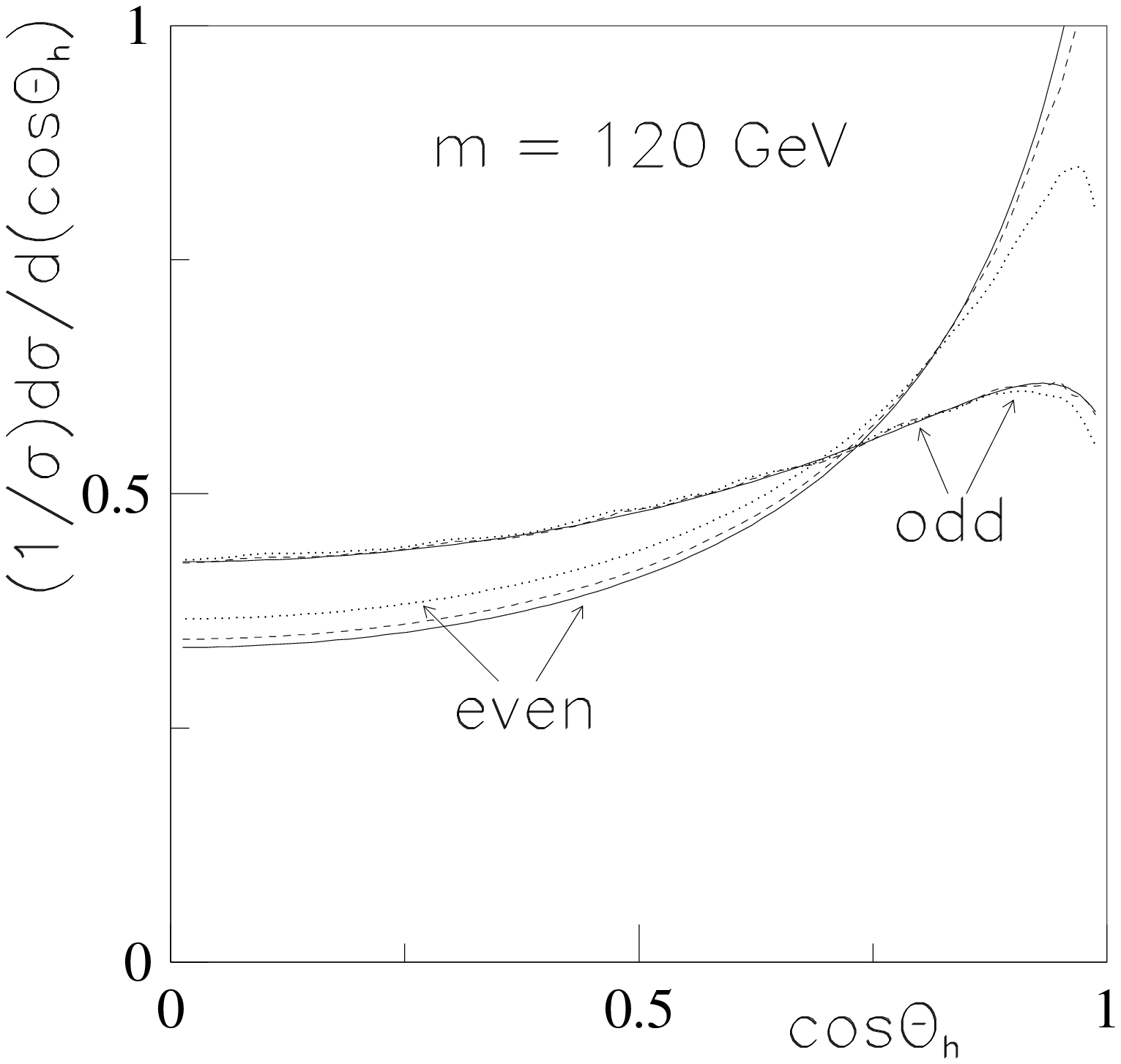}}}
\put(3.0,4.0)
{\mbox{\epsfysize=8cm\epsffile{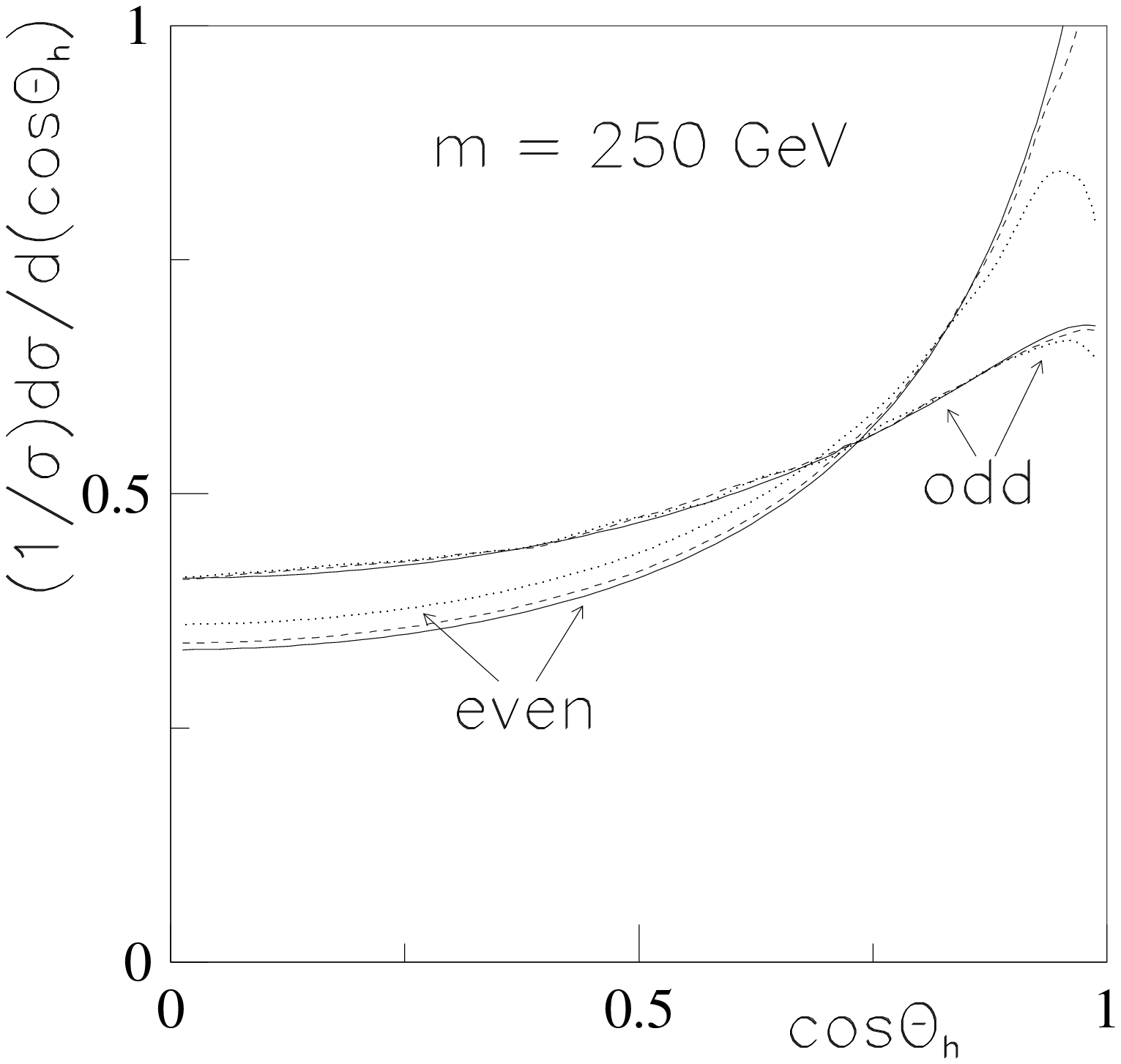}}}
\end{picture}
\end{center}
\vspace*{-4.0cm}
\caption{Distributions in $\cos\theta_h$, for $E_{\rm c.m.}=500$~GeV
and two mass values: $m_h=120$~GeV and 250~GeV.
Both the $CP$ even and the $CP$ odd cases are considered,
as indicated.
Solid curves correspond to no cuts, dashed and dotted curves
correspond to cuts on the final-state electrons at $5\deg$ and $10\deg$,
respectively.}
\end{figure}
%%%%%%%%%%%%%%%%%%%%%%%%%%%%%%%%%%%%%%%%%%%%%%%%%%%%%%%%%%%%%%%%%%%%

%%%%%%%%%%%%%%%%%%%%%%%%%%%%%%%%%%%%%%%%%%%%%%%%%%%%%%%%%%%%%%%%%%%%
\begin{figure}[htb]
\refstepcounter{figure}
\label{Fig:sigtot-mh-a1}
\addtocounter{figure}{-1}
\phantom{AAA}
\begin{center}
\setlength{\unitlength}{1cm}
\begin{picture}(16,20)
\put(3.0,5.0)
{\mbox{\epsfysize=9cm\epsffile{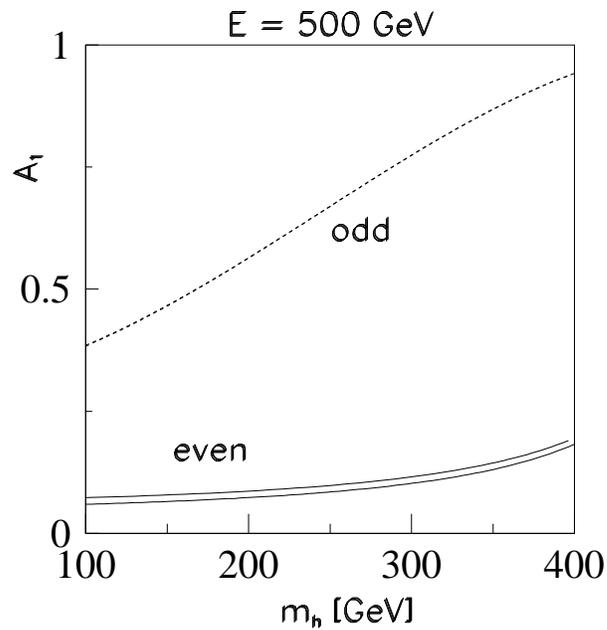}}}
\end{picture}
\end{center}
\vspace*{-3.0cm}
\caption{The bi-polarization-dependence $A_1$ 
[see Eq.~(\ref{Eq:pol-dep})] as obtained from the integrated
cross sections for Higgs production in electron-electron
collisions at $E_{\rm c.m.}=500$~GeV, for a range of Higgs masses.
Standard Model (denoted ``even'') and $CP$-odd results are shown.
For the even case, the lower curve corresponds to no cut,
whereas the upper ones are obtained with an angular cut on the
final-state electron momenta at $10\deg$.
(For the odd case, the two curves are indistinguishable.)}
\end{figure}
%%%%%%%%%%%%%%%%%%%%%%%%%%%%%%%%%%%%%%%%%%%%%%%%%%%%%%%%%%%%%%%%%%%%

%%%%%%%%%%%%%%%%%%%%%%%%%%%%%%%%%%%%%%%%%%%%%%%%%%%%%%%%%%%%%%%%%%%%
\begin{figure}[htb]
\refstepcounter{figure}
\label{Fig:phi-sig}
\addtocounter{figure}{-1}
\phantom{AAA}
\begin{center}
\setlength{\unitlength}{1cm}
\begin{picture}(16,20)
\put(0.0,11.0)
{\mbox{\epsfysize=8cm\epsffile{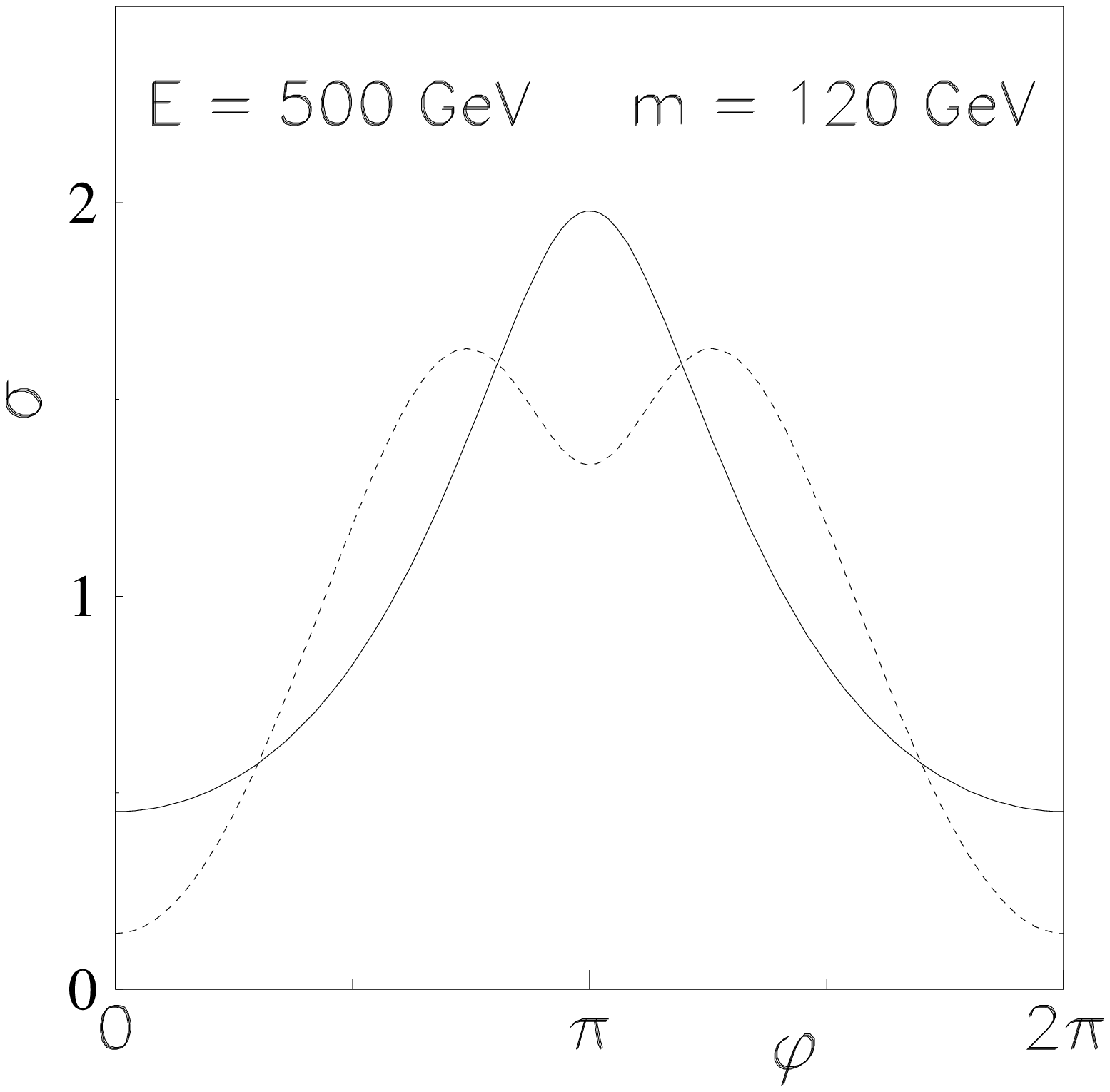}}
 \mbox{\epsfysize=8cm\epsffile{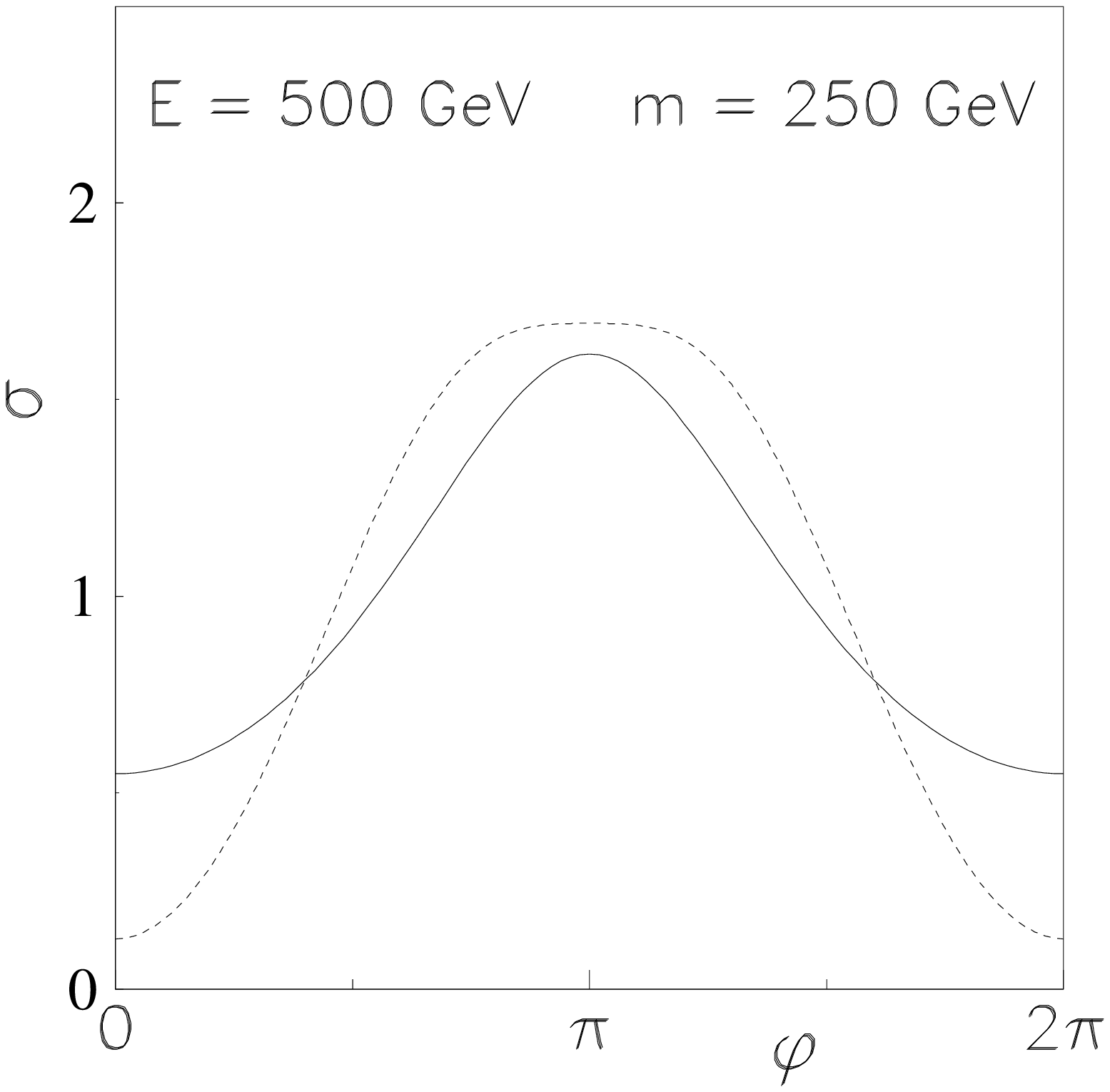}}}
\put(0.0,3.0)
{\mbox{\epsfysize=8cm\epsffile{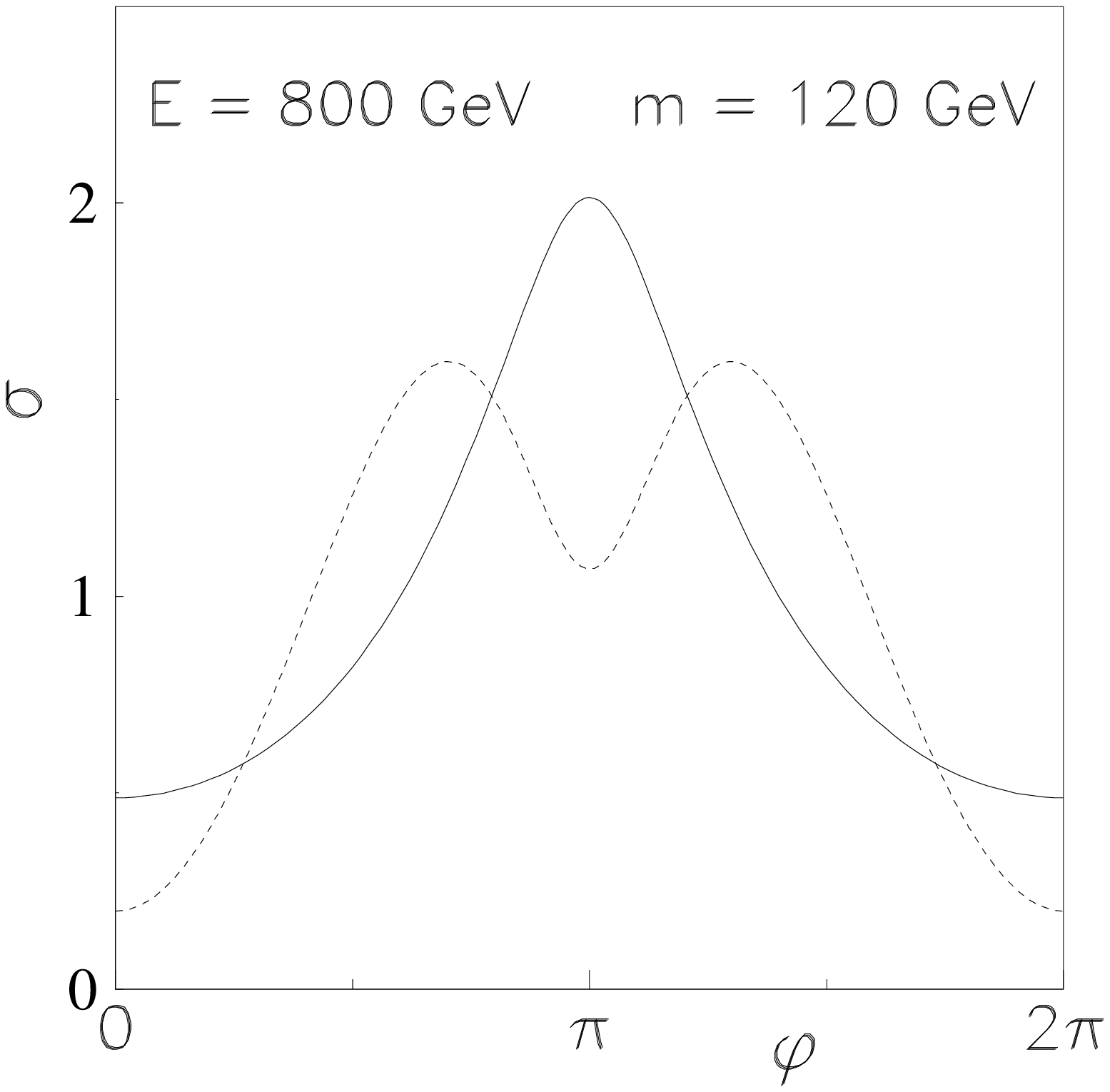}}
 \mbox{\epsfysize=8cm\epsffile{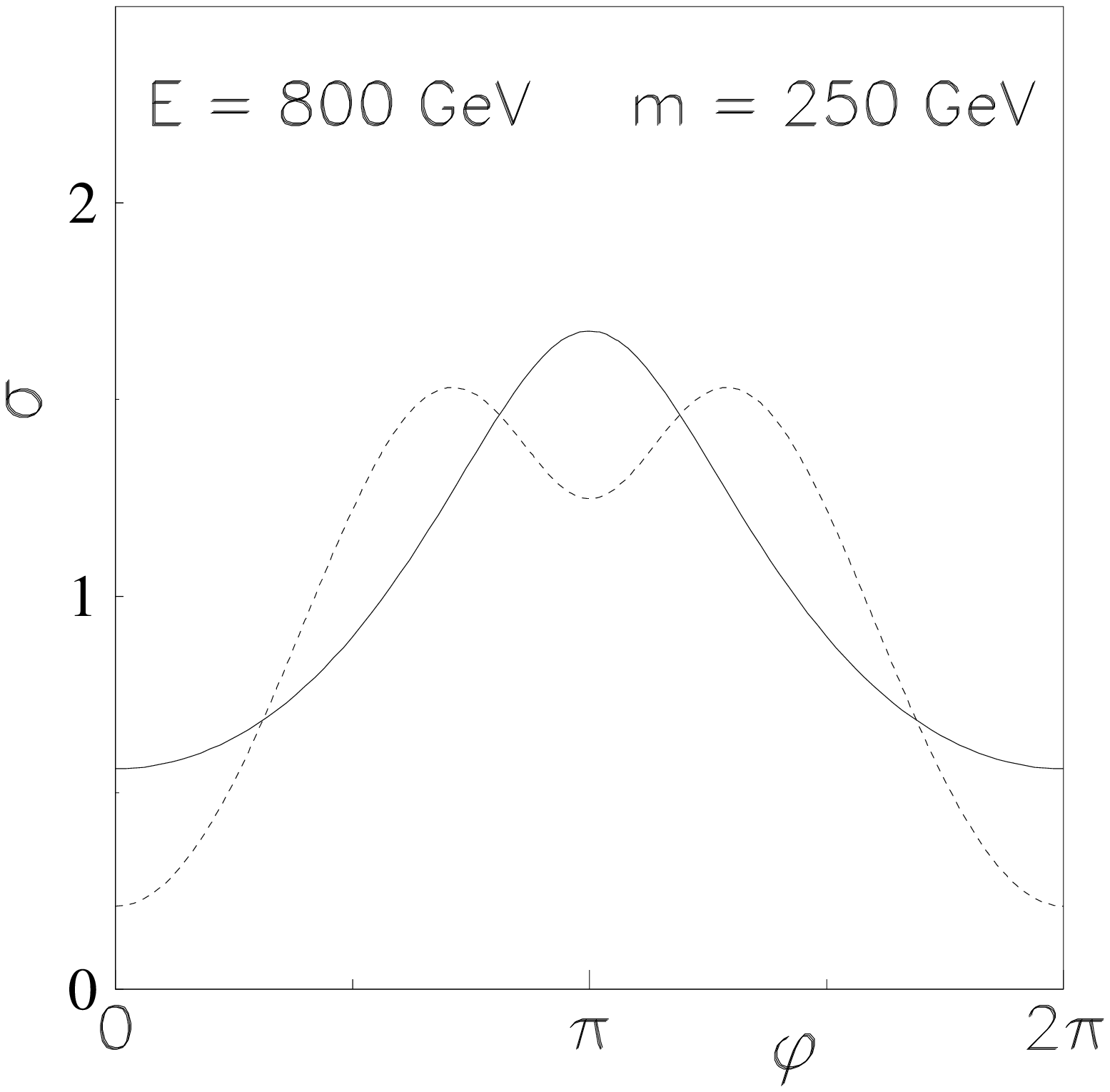}}}
\end{picture}
\end{center}
\vspace*{-3.0cm}
\caption{Azimuthal distributions at energies
$E_{\rm cm}=500$~GeV and  800~GeV,
and for Higgs masses $m_h=120$~GeV and 250~GeV, with unpolarized beams.
The polar-angle cut-off is given by $\cos\theta_{\rm c}=0.9$.}
\end{figure}
%%%%%%%%%%%%%%%%%%%%%%%%%%%%%%%%%%%%%%%%%%%%%%%%%%%%%%%%%%%%%%%%%%%%

%%%%%%%%%%%%%%%%%%%%%%%%%%%%%%%%%%%%%%%%%%%%%%%%%%%%%%%%%%%%%%%%%%%%
\begin{figure}[htb]
\refstepcounter{figure}
\label{Fig:phi-sig-500-120-pm}
\addtocounter{figure}{-1}
\phantom{AAA}
\begin{center}
\setlength{\unitlength}{1cm}
\begin{picture}(16,20)
\put(3,6.0)
{\mbox{\epsfysize=10cm\epsffile{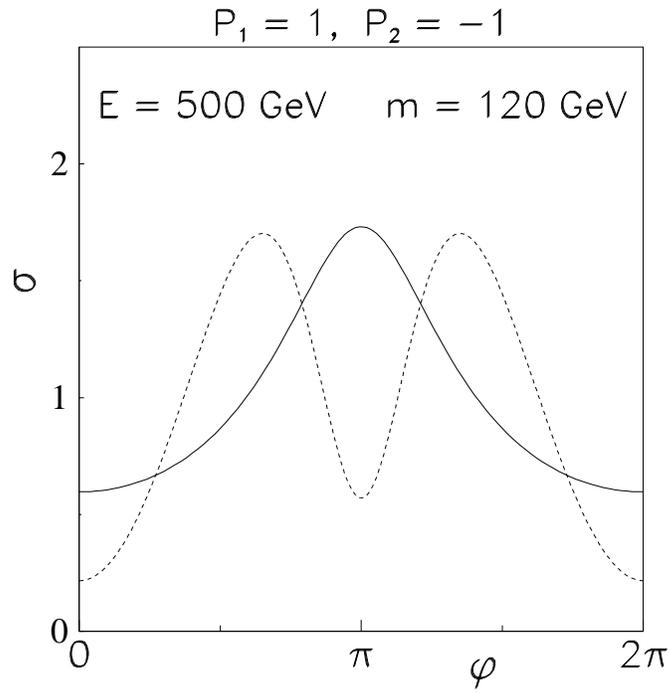}}}
\end{picture}
\end{center}
\vspace*{-4.0cm}
\caption{Azimuthal distributions at
$E_{\rm cm}=500$~GeV, $m_h=120$~GeV,
for $P_1=1$ and $P_2=-1$.
The polar-angle cut-off is given by $\cos\theta_{\rm c}=0.9$.}
\end{figure}
%%%%%%%%%%%%%%%%%%%%%%%%%%%%%%%%%%%%%%%%%%%%%%%%%%%%%%%%%%%%%%%%%%%%

%%%%%%%%%%%%%%%%%%%%%%%%%%%%%%%%%%%%%%%%%%%%%%%%%%%%%%%%%%%%%%%%%%%%
\begin{figure}[htb]
\refstepcounter{figure}
\label{Fig:cos-cos}
\addtocounter{figure}{-1}
\phantom{AAA}
\begin{center}
\setlength{\unitlength}{1cm}
\begin{picture}(16,20)
\put(-0.5,12.0)
{\mbox{\epsfysize=7cm\epsffile{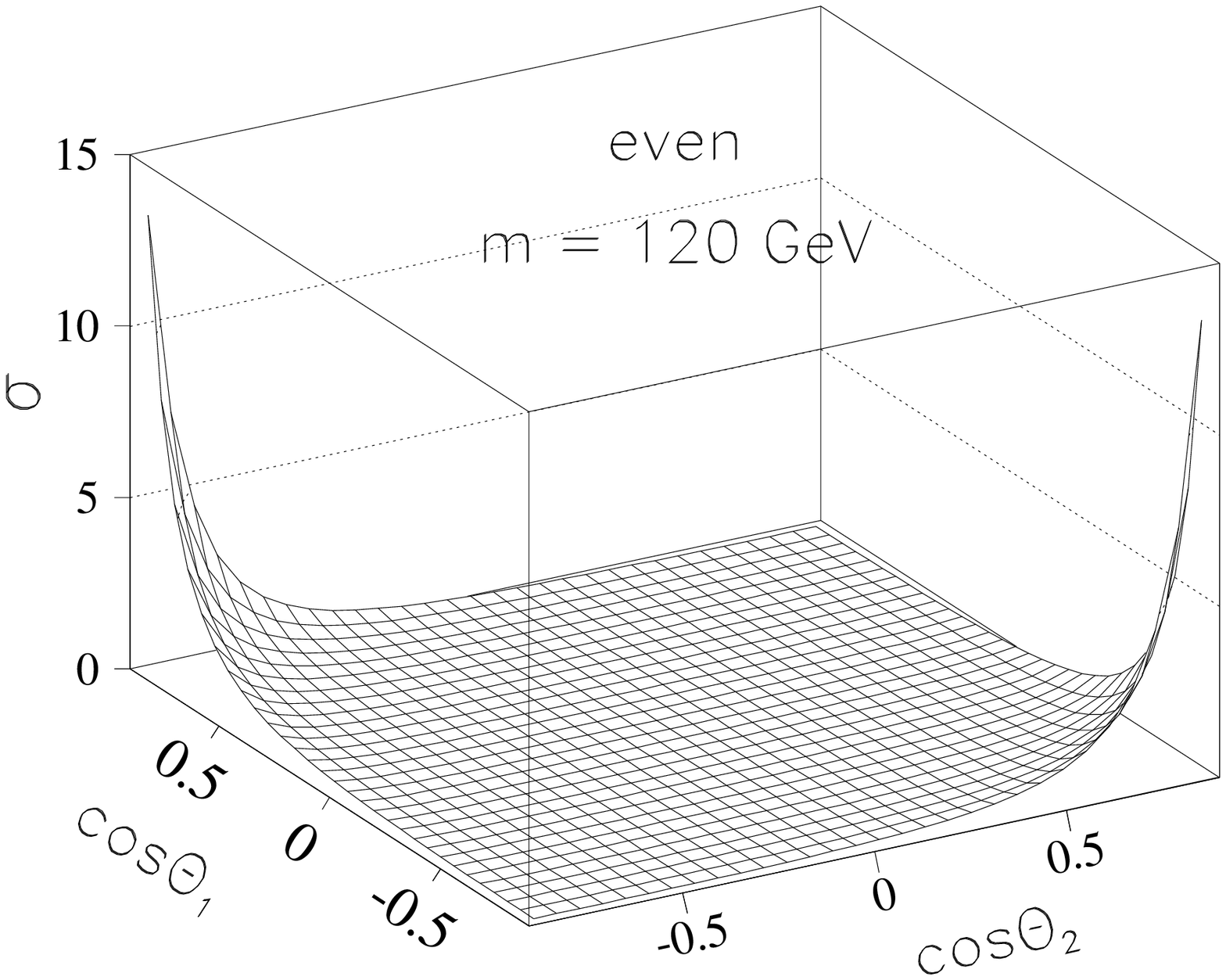}}
 \mbox{\epsfysize=7cm\epsffile{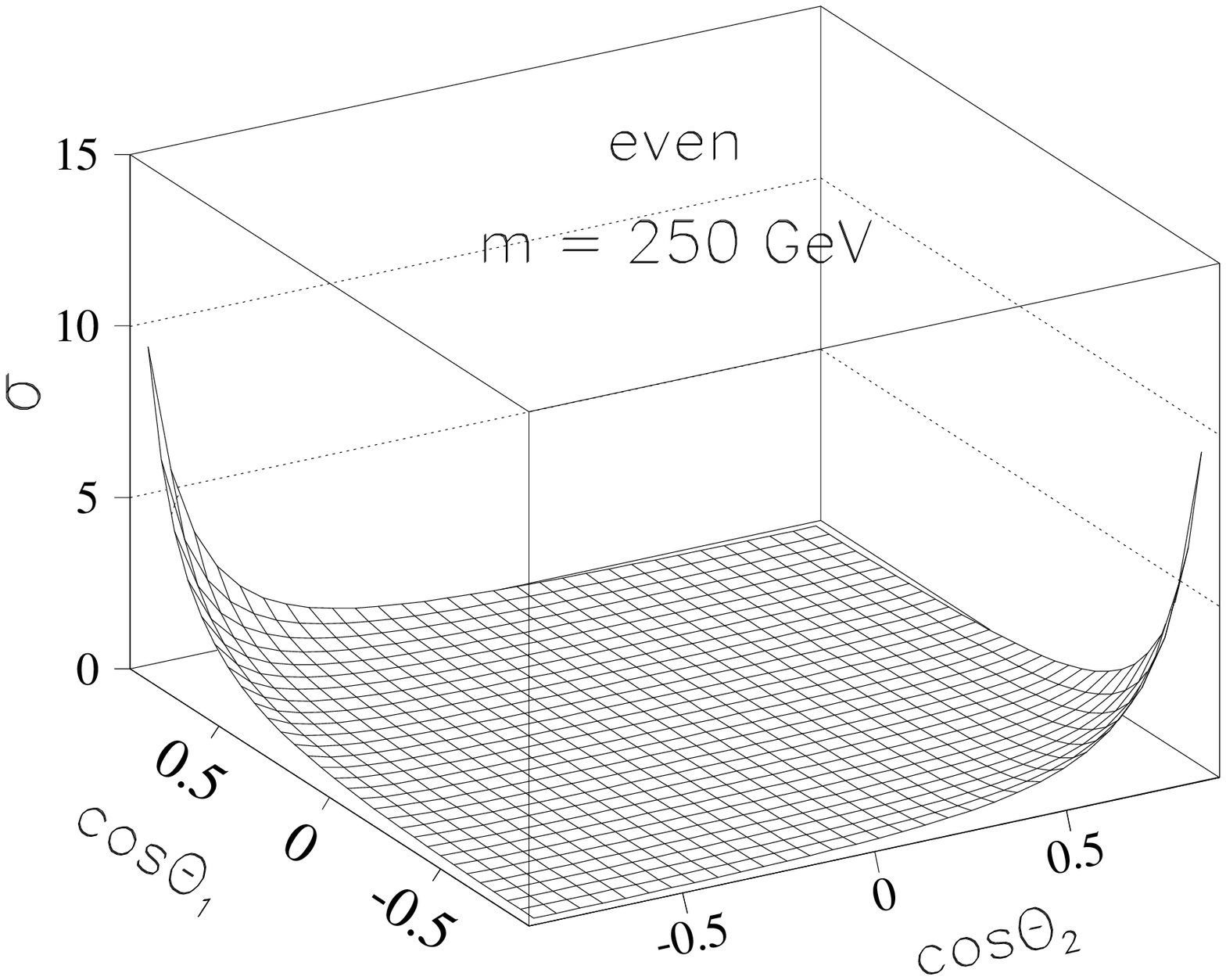}}}
\put(-0.5,4.0)
{\mbox{\epsfysize=7cm\epsffile{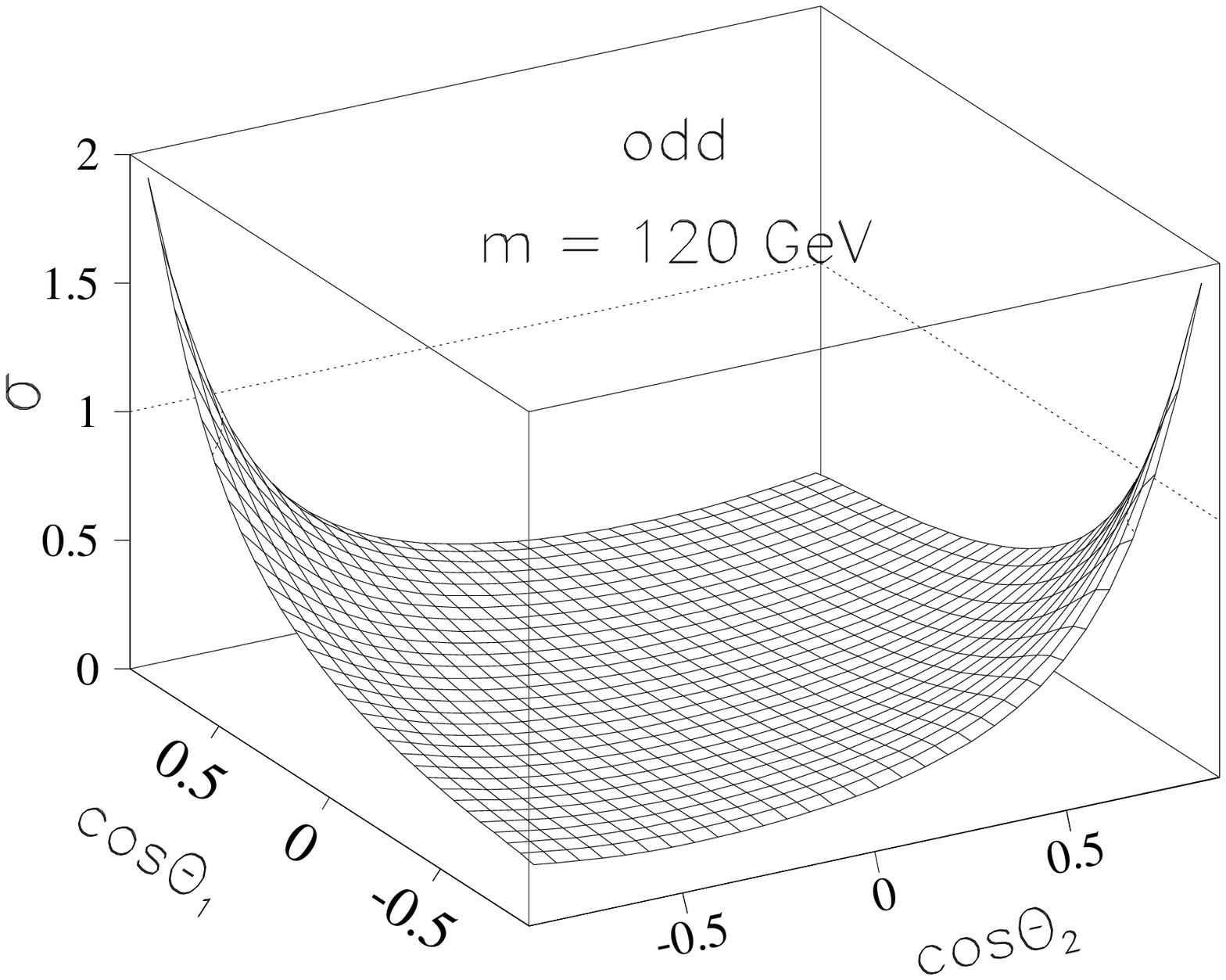}}
 \mbox{\epsfysize=7cm\epsffile{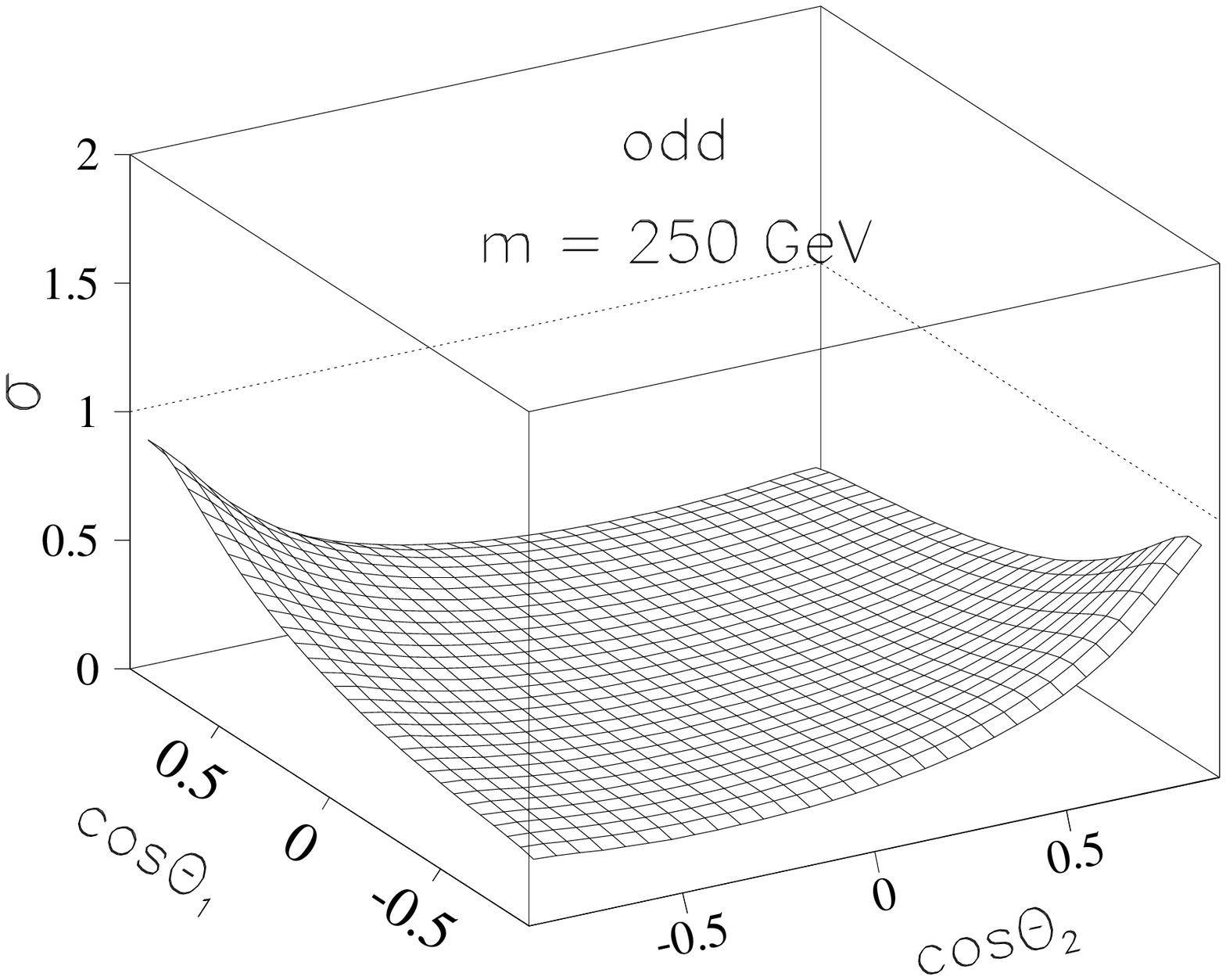}}}
\end{picture}
\end{center}
\vspace*{-4.0cm}
\caption{Normalized distributions, in the polar angles of the 
final-state electrons, $|\cos\theta_{1,2}|\le0.9$,
for $E_{\rm c.m.}=500$~GeV,
$m_h=120$~GeV, and $m_h=250$~GeV, for unpolarized beams. 
Note different scales.}
\end{figure}
%%%%%%%%%%%%%%%%%%%%%%%%%%%%%%%%%%%%%%%%%%%%%%%%%%%%%%%%%%%%%%%%%%%%

%%%%%%%%%%%%%%%%%%%%%%%%%%%%%%%%%%%%%%%%%%%%%%%%%%%%%%%%%%%%%%%%%%%%
\begin{figure}[htb]
\refstepcounter{figure}
\label{Fig:cos12-pol}
\addtocounter{figure}{-1}
\phantom{AAA}
\begin{center}
\setlength{\unitlength}{1cm}
\begin{picture}(16,20)
\put(0.0,6.0)
{\mbox{\epsfysize=8cm\epsffile{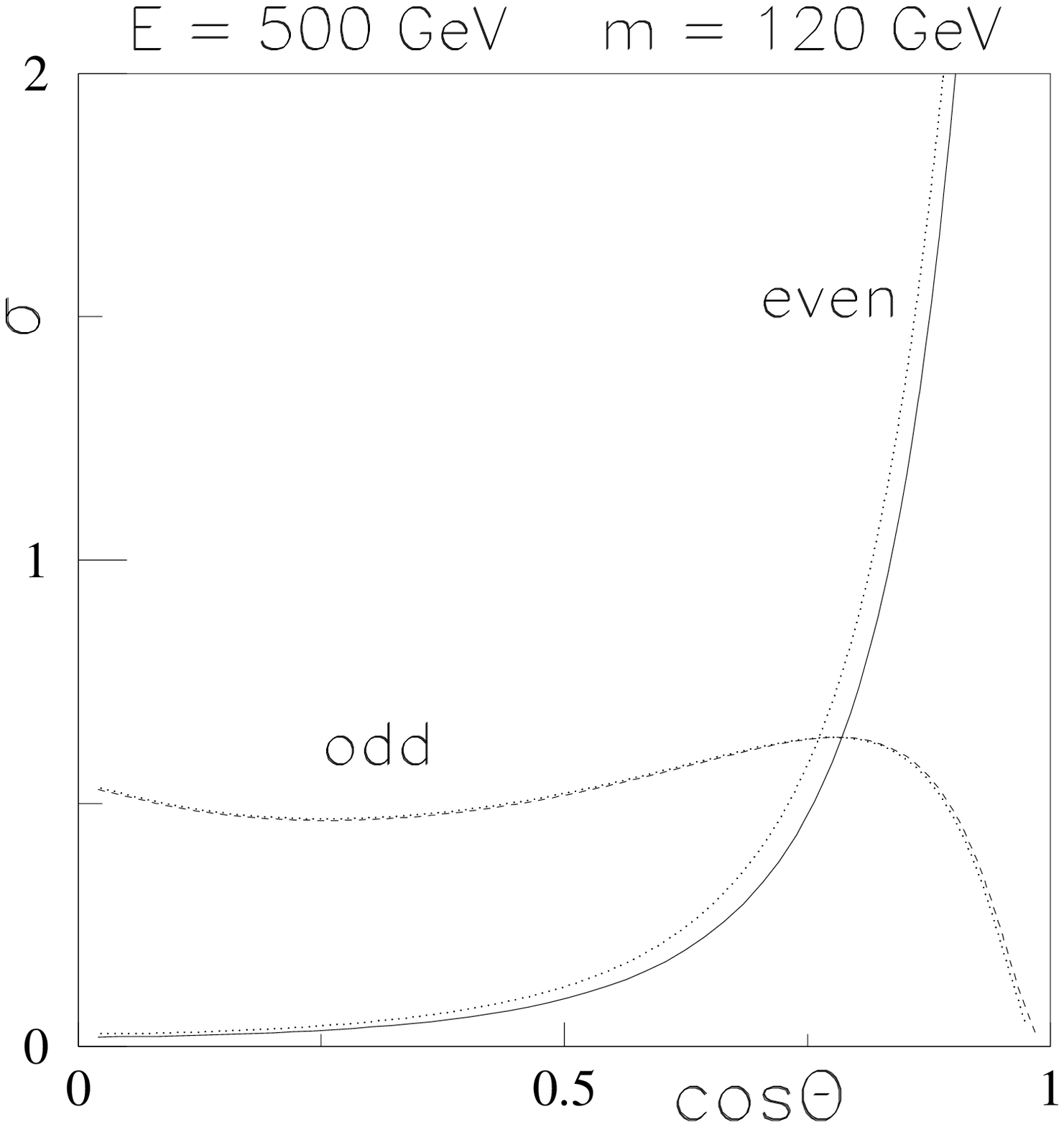}}
 \mbox{\epsfysize=8cm\epsffile{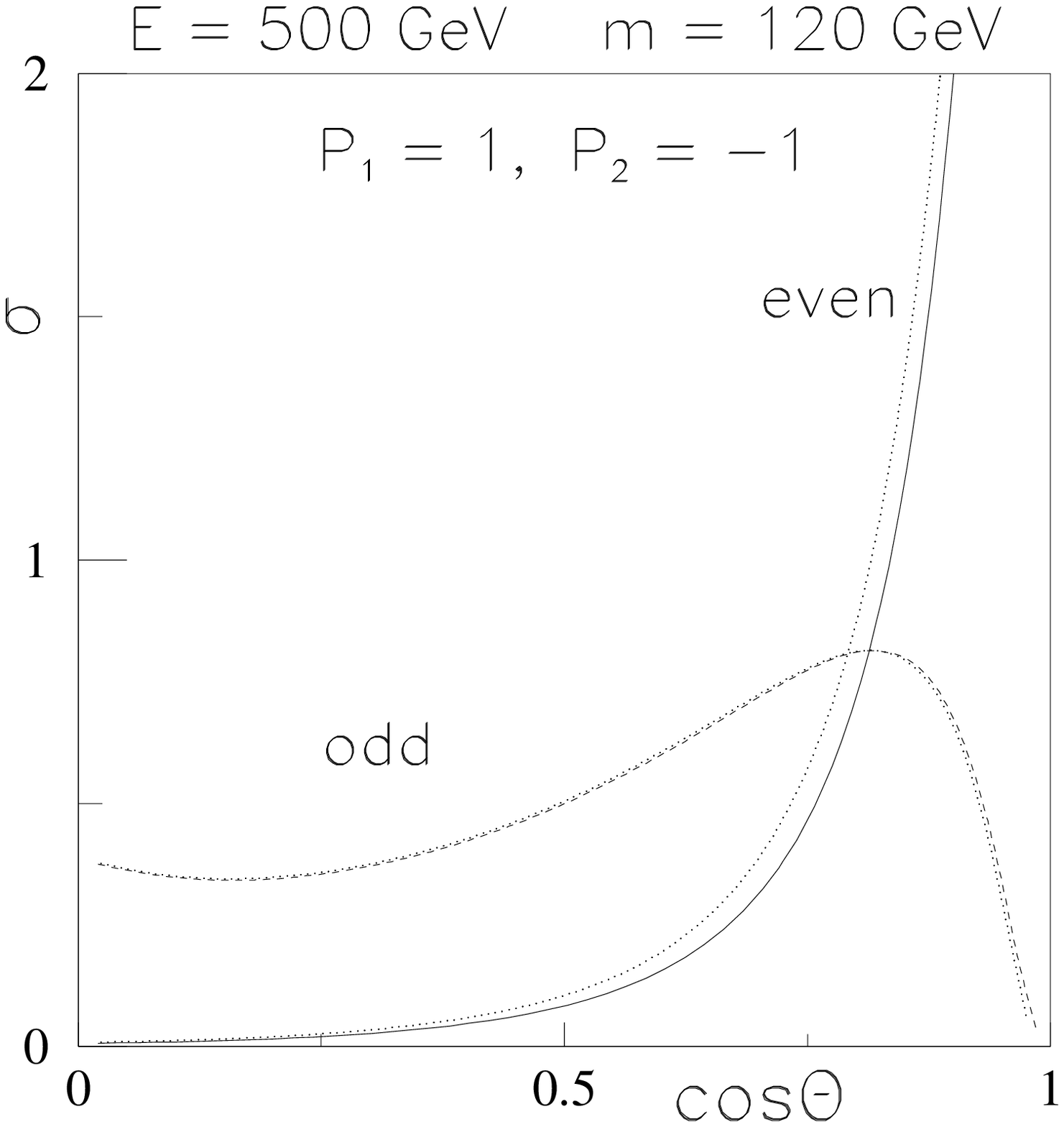}}}
\end{picture}
\end{center}
\vspace*{-4.0cm}
\caption{Distributions in $\cos\Theta$, for $E_{\rm c.m.}=500$~GeV,
$m_h=120$~GeV, and (a) unpolarized beams, (b) $P_1=1$, $P_2=-1$.
Solid and dashed: cut at $5\deg$, dotted: cut at $10\deg$.}
\end{figure}
%%%%%%%%%%%%%%%%%%%%%%%%%%%%%%%%%%%%%%%%%%%%%%%%%%%%%%%%%%%%%%%%%%%%

%%%%%%%%%%%%%%%%%%%%%%%%%%%%%%%%%%%%%%%%%%%%%%%%%%%%%%%%%%%%%%%%%%%%
\begin{figure}[htb]
\refstepcounter{figure}
\label{Fig:eps-sig-pol00}
\addtocounter{figure}{-1}
\phantom{AAA}
\begin{center}
\setlength{\unitlength}{1cm}
\begin{picture}(16,20)
\put(1.5,6.0)
{\mbox{\epsfysize=14cm\epsffile{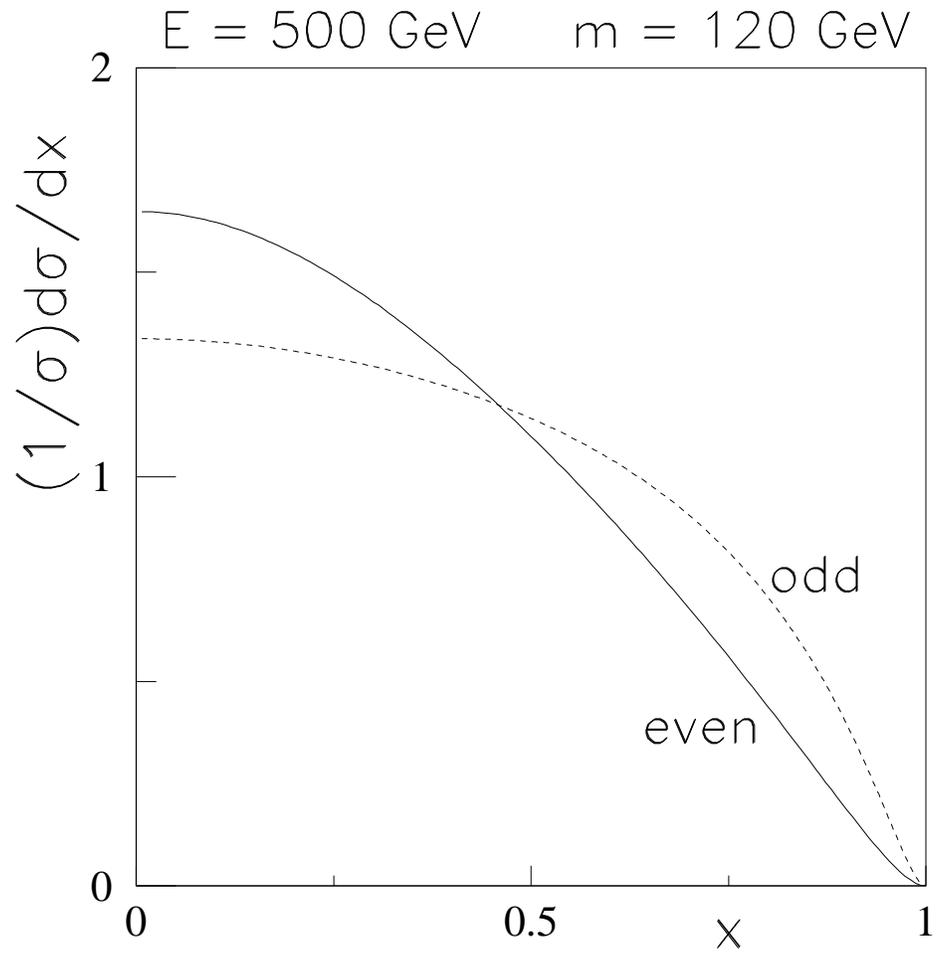}}}
\end{picture}
\end{center}
\vspace*{-4.0cm}
\caption{Distributions in relative final-state electron energy,
$x=\varepsilon/\varepsilonmax$, 
for $E_{\rm c.m.}=500$~GeV,
$m_h=120$~GeV, and unpolarized beams.}
\end{figure}
%%%%%%%%%%%%%%%%%%%%%%%%%%%%%%%%%%%%%%%%%%%%%%%%%%%%%%%%%%%%%%%%%%%%

%%%%%%%%%%%%%%%%%%%%%%%%%%%%%%%%%%%%%%%%%%%%%%%%%%%%%%%%%%%%%%%%%%%%
\begin{figure}[htb]
\refstepcounter{figure}
\label{Fig:phi-sig-cp}
\addtocounter{figure}{-1}
\phantom{AAA}
\begin{center}
\setlength{\unitlength}{1cm}
\begin{picture}(16,20)
\put(3,6.0)
{\mbox{\epsfysize=10cm\epsffile{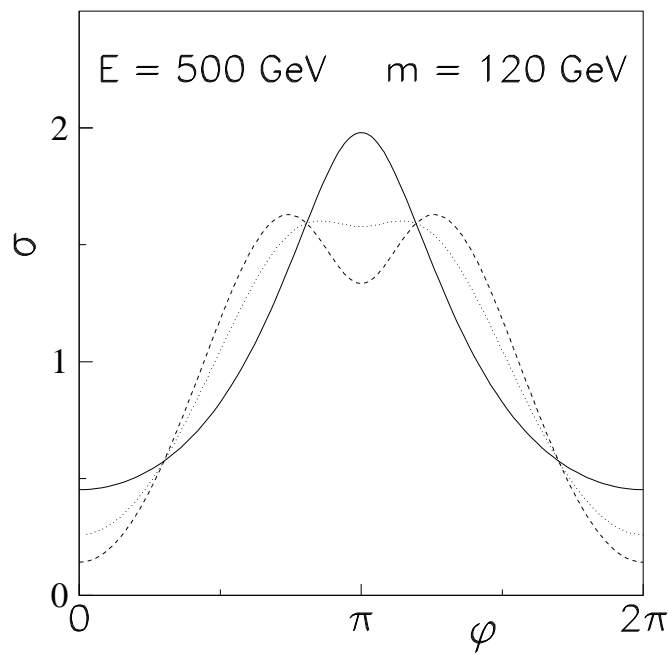}}}
\end{picture}
\end{center}
\vspace*{-4.0cm}
\caption{Azimuthal distributions for 
$E_{\rm cm}=500$~GeV, $m_h=120$~GeV.
Solid: $CP=+1$, dashed: $CP=-1$, dotted: $CP$ violated, with $\eta=0.5$.
Polar-angle cuts: $|\cos\theta_c|\le0.9$}

\end{figure}
%%%%%%%%%%%%%%%%%%%%%%%%%%%%%%%%%%%%%%%%%%%%%%%%%%%%%%%%%%%%%%%%%%%%

%%%%%%%%%%%%%%%%%%%%%%%%%%%%%%%%%%%%%%%%%%%%%%%%%%%%%%%%%%%%%%%%%%%%
\begin{figure}[htb]
\refstepcounter{figure}
\label{Fig:phi-sig-cp-as}
\addtocounter{figure}{-1}
\phantom{AAA}
\begin{center}
\setlength{\unitlength}{1cm}
\begin{picture}(16,20)
\put(3,6.0)
{\mbox{\epsfysize=10cm\epsffile{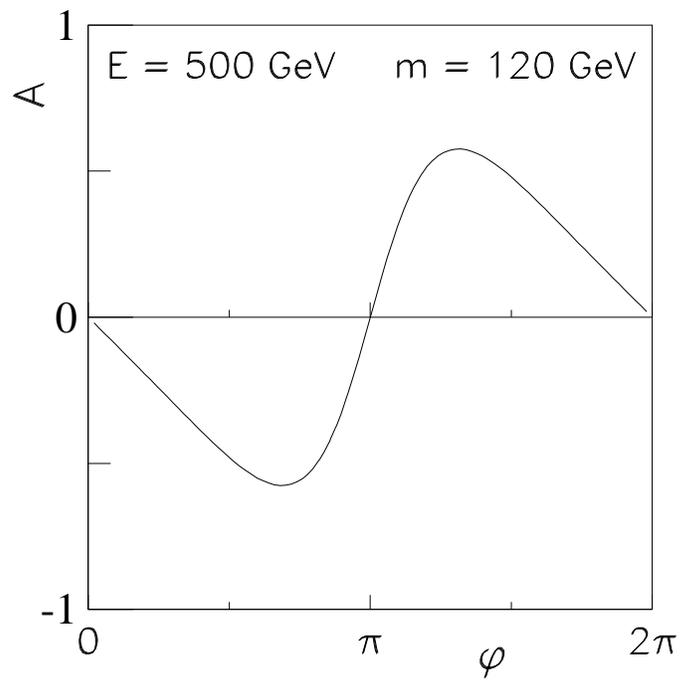}}}
\end{picture}
\end{center}
\vspace*{-4.0cm}
\caption{$CP$-violating asymmetry $A$ of Eq.~(\ref{Eq:CP-A}), for
$E_{\rm cm}=500$~GeV, $m_h=120$~GeV, and $\eta=0.5$.
Polar-angle cuts: $|\cos\theta_c|\le0.9$}

\end{figure}
%%%%%%%%%%%%%%%%%%%%%%%%%%%%%%%%%%%%%%%%%%%%%%%%%%%%%%%%%%%%%%%%%%%%

\end{document}